\documentclass[prd,aps,letterpaper,floatfix,superscriptaddress,preprintnumbers,twocolumn,10pt,nofootinbib]{revtex4-1}

\usepackage{geometry}
\usepackage{hyperref}
\usepackage{ulem}
\hypersetup{
    colorlinks=true,
    linkcolor=blue,
    filecolor=magenta,      
    urlcolor=cyan,
}
\usepackage{amsmath,amssymb,graphicx}
\graphicspath{{figures/}}

\newcommand{\be}{\begin{equation}}
\newcommand{\ee}{\end{equation}}

\usepackage{tikz}
\usetikzlibrary{shapes,arrows}
\tikzstyle{decision} = [diamond, draw, fill=green!30, 
    text width=5em, text badly centered, node distance=2cm, inner sep=0pt]
\tikzstyle{block} = [rectangle, draw, fill=blue!30, 
    text width=6em, text centered, rounded corners, minimum height=4em]
\tikzstyle{line} = [draw, -latex']
\tikzstyle{cloud} = [draw, ellipse,fill=red!30, node distance=2cm,
    minimum height=2em]
\tikzstyle{process} = [rectangle, minimum width=3cm, minimum height=1cm, text centered, text width=3cm, draw=black, fill=orange!30]   

\usepackage{ulem}
\begin{document}

\title{Higgs boson tagging with the Lund jet plane}
\author{Charanjit K. Khosa} 
\email{charanjit.kaur@ge.infn.it}
\affiliation{Dipartimento di Fisica, Universit\`a di Genova and INFN, Sezione di Genova, Via Dodecaneso 33, 16146, Italy}
\author{Simone Marzani}
\email{simone.marzani@ge.infn.it }
\affiliation{Dipartimento di Fisica, Universit\`a di Genova and INFN, Sezione di Genova, Via Dodecaneso 33, 16146, Italy}

\begin{abstract}

We construct a procedure to separate boosted Higgs bosons decaying into hadrons, from the background due to strong interactions. 
We employ the Lund jet plane to obtain a theoretically well-motivated representation of the jets of interest and we use the resulting images as the input to a convolutional neural network classifier. 
 In particular, we consider two different decay modes of the Higgs boson, namely into a pair of bottom quarks or into light jets, against the respective backgrounds. For each case, we consider both a moderate- and high- boost scenario. 
 The performance of the tagger is compared to what is achieved using a traditional single-variable analysis which exploits a QCD inspired color-singlet tagger, namely the jet color ring observable. 
    
\end{abstract}
\maketitle
\section{Introduction}

High-energy collision events at the CERN Large Hadron Collider (LHC) are characterised by copious hadronic activity. Not only protons are strongly-interacting, but also elementary particles produced in their collisions often carry color charges, resulting into final-states characterized by many hadrons.
One powerful way to deal with the complex environment of hadronic activity at the LHC is to note that final-state hadrons tend to be produced in fairly collimated sprays, along directions that we can think of as being the ones of the originating hard (i.e.\ with large transverse momentum) particles. These collimated sprays of hadrons are called jets.

The vast majority of high transverse-momentum ($p_T$) jets are QCD jets, i.e.\ they originate from the fragmentation of a high-$p_T$ parton (a quark or a gluon). However, every particle that decays hadronically can, if sufficiently boosted, give rise to a collimated spray of hadrons, which is reconstructed as a jet. Therefore, one key aspect in the context of LHC phenomenology is to correctly identify the nature of the particles originating these jets. This endavor is often referred to as jet tagging.  
For example, top quarks, electroweak gauge bosons, and the Higgs boson, if produced at high-$p_T$, can be identified using jet tagging techniques. These algorithms can also be applied in searches for new physics.  Therefore, every tiny gain in jet tagging efficiency is important for both for measurements and new physics searches at the LHC. 
Tremendous progress over the past decade (see e.g.~\cite{Marzani:2019hun}) has led to the development of jet substructure techniques, including tagging algorithms, that are very efficient in distinguishing signal jets from background ones, where the latter ones are often QCD jets. Furthermore, the application of field-theory methods to jet physics has provided us with a deeper understanding of jet substructure. This, in turns, has allowed us to develop algorithms that are not only efficient, but robust. 

In the last few years a complementary approach to jet substructure has risen to prominence. 
The rapid development of machine-learning (ML) techniques is deeply changing the way particle physics analyses are conducted.
In recent years, many groups have been exploring the potential of ML in the context of LHC phenomenology.
There have been several suggestions, ranging from top tagging~\cite{Kasieczka:2019dbj}, constraining Wilson coefficients of higher dimensional operators in effective field theory (EFT) framework~\cite{Freitas:2019hbk,Brehmer:2018kdj,Brehmer:2018eca,Chen:2020mev}, quark-gluon tagging~\cite{Kasieczka:2018lwf}, model agnostic new physics searches~\cite{Kasieczka:2021xcg}, jet substructure~\cite{Kasieczka:2020nyd}, likelihood encoding~\cite{Coccaro:2019lgs} and many others.  In addition, there is a continuous effort to adapt these techniques for the various steps of data analysis i.e.\ trigger, event reconstruction, particle identification, heavy flavour tagging, jet tagging, and signal and background classification (for recent reviews see e.g.~\cite{Radovic:2018dip, Guest:2018yhq}). 

In the context of jet substructure, classification algorithms that exploit deep neural networks (NN) have been shown to often out-perform the more traditional tagging techniques, which indeed can be thought of as lower-dimensional projections of the complex sets of inputs that are fed to the NN (see e.g.~\cite{Larkoski:2017jix}).

One intriguing aspect of ML techniques is their ability to make data-driven decisions without using the prior knowledge of the underlying theory. This has sparked an interesting debate on whether one should take a rather agnostic approach and favor raw data, such as particles' kinematics, as inputs to NN, or whether one should make good use of the expert-knowledge developed thanks to our theoretical understanding of the underlying physical processes, and therefore exploit higher-level, theory-inspired, objects as inputs to the NN~\footnote{We note that a comparison between different approaches has been recently performed in the context of top-tagging~\cite{Kasieczka:2019dbj}.}.
In this context, particle physics in general, and jet physics in particular, find themselves in a rather unique position to address these types of questions, because, thanks to the Standard Model, we have a deep understanding of the physical processes we are studying. For instance, we can make use of this knowledge to better understand what kind of information ML classifiers are exploiting~\cite{Kasieczka:2020nyd,Larkoski:2019nwj}.
Therefore a growing number of studies that combine the power of ML classification with theory-inspired variables has appeared in recent times. These include the use of $N$-subjettiness variables~\cite{Datta:2017rhs,Aguilar-Saavedra:2020sxp, Aguilar-Saavedra:2020uhm}, energy flow variables~\cite{Komiske:2017aww, Komiske:2018cqr}, two-point energy correlations~\cite{Chakraborty:2020yfc}, and jet charge~\cite{Fraser:2018ieu} for multi-prong jet tagging. 

A very interesting observable, both in the context of boosted object tagging as well as Standard Model measurements is the Lund jet plane~\cite{Dreyer:2018nbf}. Similarly to calorimeter jet images~\cite{deOliveira:2015xxd, Komiske:2016rsd}, the (primary) Lund jet plane is a two-dimensional representation of a jet. However, differently to the calorimetric approach, which constructs images in the pseudorapidity-azimuth ($\eta,\phi$) plane, the Lund jet plane image is given in terms of kinematic variables that better describe the emissions in the jet. 
By construction, the Lund jet plane allows one to separate different physical effects. Indeed, the Lund plane was originally designed to describe the way phase-space is filled by Monte Carlo parton-showers. It is also often used in the context of analytic resummation. Recently, the all-order structure of the Lund jet plane density has also been computed~\cite{Lifson:2020gua}.
The Lund jet plane has been used together with Long Short-Term Memory (LSTM) networks and, more recently, with graph neural networks for $W$ boson and top tagging~\cite{Dreyer:2018nbf,Dreyer:2020brq}. It is also a plausible choice for event generation using generative models~\cite{Carrazza:2019cnt}. Lund basis for the jet declustering history is also used in the context of unsupervised new physics search method~\cite{Dillon:2020quc}. 
From the experimental side, the Lund jet plane density has been measured by the ATLAS~\cite{Aad:2020zcn} and ALICE~\cite{ALICE-PUBLIC-2021-002} collaborations.

In the following, we will study the performance of the Lund jet plane in the context of boosted Higgs boson identification.
We shall consider both the decay of the Higgs boson in a pair bottom ($b$) quarks that subsequently fragment into two $b$-(sub)jets, henceforth $H\to b\bar b$, and the decay of the Higgs boson into light (sub)jets, henceforth $H \to gg$.

The Lund jet plane offers us the opportunity to address interesting theoretical questions and simultaneously provides good classification performance. 
In particular, we are going to use the primary Lund plane, which is a proxy for the two-dimensional phase space of the leading emission in the jet. 
Resulting Lund jet plane images for signal and background are used as inputs to a convolutional neural network (CNN). 
CNNs are known to be very efficient for image data sets. However, they have been mostly tested in cases where images are constructed, unlike our case, from raw data or, in other words, unprocessed data set.  Within the high-energy physics community, CNNs are showing exciting potential for instance, in the context of top tagging~\cite{CNNKasieczka, CNNsShih}, dark matter searches (see e.g.~\cite{Khosa:2019qgp,Khosa:2019kxd}), disentangling Higgs production modes~\cite{Chung:2020ysf}, and anomaly detection~\cite{Khosa:2020qrz}.

We believe that this study is an interesting addition to the rather extensive literature on ML-based Higgs taggers. Some of these methods exploit low-level inputs to construct jet images and use CNN~\cite{Lin:2018cin, Alves:2019ppy, Guo:2020vvt, Li:2020grn} or interaction networks~\cite{Moreno:2019neq}. Other approaches combine the use of NN with $N$-subjettiness and related variables~\cite{Datta:2017lxt,Datta:2019ndh}. 
The potential of ML techniques has also been explored for a better reconstruction of the Higgs (decaying to $b\bar b$) for the trilinear coupling measurements ~\cite{Amacker:2020bmn,Abdughani:2020xfo}. Other ML based Higgs studies involve, for instance, disentangling Higgs production modes~\cite{Chung:2020ysf}, invisible Higgs decay~\cite{Ngairangbam:2020ksz}, Higgs width measurement\cite{Harris:2019qwx}, Higgs cascade decay~\cite{Englert:2020ntw}, bottom Yukawa couplings~\cite{Grojean:2020ech}.

The primary aim of this work is to explore the Lund jet plane tagging performance for the boosted Higgs boson decays $H\to b\bar b$ and $H\to gg$. The main backgrounds to these processes are given, respectively, by high transverse-momentum jets that are doubly $b$-tagged, mostly driven by $g\to b\bar b$ splittings, or by generic light jet production ($jj$). Therefore, we can view Higgs tagging as an example of a more general problem that aims to identify color-singlet states that are decaying hadronically from states that belong to other representations of the $SU(3)$ color group.
 Recently, the jet color ring~\cite{Buckley:2020kdp} observable was proposed for color-singlet identification.
 This observable was derived using ratios of signal and background matrix elements, in the soft limit. 
 By construction, the jet color ring is monotonic with respect to the likelihood ratio and therefore, it represents an optimal tagger, given the approximations made in its derivation.
 Indeed, it was found~\cite{Buckley:2020kdp} that this simple observable offers good tagging efficiency for $H \to b \bar{b}$. However, it fails miserably when considering the $H\to gg$ case. 
 In this study, we compare the tagging efficiency obtained with the Lund jet plane to the jet color ring, with the aim of identifying better ways to harvest the information on color radiation, in order to build efficient taggers.

This paper is organized as follows. In section~\ref{set-up}, we describe the main building blocks of this work including the analysis set-up. Section~\ref{hbb} and~\ref{hgg} are devoted to the $H \to b \bar b$ and the $H \to gg$ analyses, respectively. We perform our studies in two different kinematic region: moderate-boost ($p_T>250$~GeV) and high-boost ($p_T>550$~GeV). 
We summarise our results in the last section.

\section{Analysis Strategy\label{set-up}}
Before discussing the details of our analyses, we describe the two main theoretical building blocks of the work, i.e.\ the Lund jet plane and the jet color ring.
\subsection{Lund jet plane}
The idea of constructing a Lund plane for an individual jet was proposed a couple of years ago in Ref.~\cite{Dreyer:2018nbf}. That work sparked new interest in the use of the Lund plane in jet physics, beyond its traditional application for the description of the emissions' phase space, primarily in the context of parton showers and resummation.

The Lund jet plane is a physically-motivated, QCD-driven, representation of a jet.
It is formed parsing backwards the Cambridge-Aachen (C/A)~\cite{Dokshitzer:1997in,Wobisch:1998wt} clustering history of the jet. The procedure starts by undoing the final clustering step and by recording the kinematics of the splitting. The primary Lund jet plane is obtained by iterating the above procedure, always following the hardest branch in each splitting. The most useful representation of the recorded information is given in terms of a double-logarithmic plane. Following~\cite{Dreyer:2018nbf}, we choose as variables the azimuth-rapidity separation of the branches involved in the splitting and the transverse momentum of the {\it emission} with respect to the {\it emitter}~\footnote{Given the splitting $p\to p_a \, p_b$, with $p_{Ta}>p_{Tb}$, we consider $p_b$ to be the emission and $p$ the emitter, while $\Delta=\sqrt{(\phi_a-\phi_b)^2+(y_a-y_b)^2}$.}. 
Although other representations are possible, the $(\ln \frac{1}{\Delta}, \ln \frac{k_t}{\text{GeV}})$ plane has the advantage of a clear separation between different physical effects, such as collinear ($\Delta\simeq 0$) from large-angle ($\Delta \simeq 1$) emissions, and perturbative ($k_t\gg1$~GeV) from non-perturbative ($k_t \lesssim 1$~GeV).

\subsection{Jet color ring}
The jet color ring ($\mathcal{O}$)~\cite{Buckley:2020kdp} is a color-singlet tagger for boosted two-prong decays. It is defined as 
\be \mathcal{O} = \frac{\Delta_{ka}^2+\Delta_{kb}^2}{\Delta_{ab}^2}\ee
where $a$ and $b$ are primary subjets (i.e.\ the leading subjets or the subjets that have been tagged according to the decay's properties, e.g. \ $b$-tagged), while $k$ is leading remaining subjet. This third subjet is taken as a proxy for soft-gluon emission in the jet. As before, $\Delta$ measures the separation, in the rapidity-azimuth plane, between the subjet pairs. 
Color conservation dictates that $a$ and $b$ are color-connected if the decaying state is a color-singlet. In such case, $k$ will be predominantly emitted in between the legs of the $ab$ dipole and, as a result, the distribution of the color ring will be peaked at small $\mathcal{O}$.

Both the Lund jet plane and the jet color ring receive their inspiration from a first-principle analysis of the physical process we are interested in. 
However, it is clear that the Lund jet plane provides us with more information than the jet color ring. The Lund jet plane is a two-dimensional representation of the jet of interest, in which each splitting is mapped into a point of the $(\ln \frac{1}{\Delta}, \ln \frac{k_t}{\text{GeV}})$ plane. On the other hand, the color ring is a more standard, and much simpler, jet observable, which associates a jet to a single value of $\mathcal{O}$.
In the following, we will compare the tagging performance of the Lund jet plane, used as the input to a CNN, and of a much simpler tagging strategy based on just the jet color ring. On the one hand, we are interested in exploring possible gain brought by the Lund jet plane and the use of ML in the $H\to b\bar b$ case, where we know we can already achieve good performance using the jet color ring.
On the other hand, we also want to study the $H\to gg$ case, where the color ring offers essentially no discrimination power.

\subsection{Simulation set-up}
We now discuss the methodology of our work. First, we discuss the event generation set-up, then the analysis cuts and the steps to construct our observables of interest. Finally, the generic features of the CNN architecture are described.

\subsubsection{Event generation}
We use \textsc{Madgraph 2.7.2}~\cite{Alwall:2014hca} to generate the events for signal and background processes at $\sqrt{s}=14$ TeV for both the $H \to b \bar b$ and $H \to g g$ analyses. The considered signal process is $p p$ $\rightarrow$ $ZH$ where $Z \rightarrow \mu^+ \mu^-$ and $H \rightarrow$ $b \bar b$ or $gg$. Background processes are $Zbb$ and $Zjj$ for the $H \to b \bar b$ and $H \to g g$ analyses, respectively. A generation-level $p_T^{\mu\mu}$ cut of 200 and 500 GeV is imposed for the moderate- and high-boost scenarios, respectively. Further cuts on pseudo-rapidity of leptons ($|\eta_l|<$2.5) and jets ($|\eta_j|<$5.0) are imposed. \textsc{Pythia 8}~\cite{Sjostrand:2006za, Sjostrand:2007gs} is used to simulate the parton-shower and the hadronization process. We consider the particles with $|\eta|<5$ to form jets.

We cluster jets using the anti-$k_t$ algorithm~\cite{antikt} with $R=1.0$, using its implementation in \textsc{Fastjet 3.3.3}~\cite{fastjet}. The large jet radius should ensure that the decay products of Higgs are reconstructed in a single jet, in both boosted scenarios. Hard muons from the $Z$ boson decay are excluded from the clustering.
We further check jet-lepton separation. 
The leading jet of $p_T> 250$ GeV ($p_T> 550$~GeV for the high-boost benchmark) is considered for further analysis. 
Following standard practice, jets with an invariant mass close to the Higgs mass are considered. In particular, we keep signal and background jets with (110$<m_J<$140 GeV). 
The Lund jet plane and the jet color ring are then measured on the leading jet, as will be detailed below.

We note that the analysis efficiency is different for the signal and background processes as well as for the different $p_T$ values considered here. Consequently, the total number of events generated differs in all the cases and a large enough samples are generated in order to obtain at least 100K events for each case.

\subsubsection{Constructing the jet color ring}
In order to construct the jet color ring, we further need to identify subjets within the leading jet.  

Following Ref.~\cite{Buckley:2020kdp}, we consider charged particles with $p_T>$ 500 MeV and $|\eta|<5$, and construct $R=0.2$ track jets using the anti-$k_t$ algorithm. Track-jets with $p_T>5$~GeV and $\Delta<0.8$ with respect to the leading jet are considered as inputs for the jet color ring. For $H \to b \bar b$ analysis, these track-jets are further identified as $b$-jets or light-jets. %
Here, we employ a very crude approximation for the $b$-tagging procedure. We use 
the truth information of $b$-partons and calculate the $\Delta$ separation of track jets with $b$-partons. If $\Delta_{jb}<0.2$ or $\Delta_{j\bar b}<0.2$ then we $b$-tag them. 

\subsubsection{Mapping events to the Lund jet plane}
We construct the primary Lund plane of the leading jet for the same events that pass the selection cuts described for the jet color ring. 
First, we re-cluster the leading jet using the C/A algorithm with the maximum allowed jet radius. The Lund generator module of the \textsc{FastJet contrib} code is used to get the Lund coordinates of the declustering of C/A jet. 
The primary plane images are created using $\ln (1/\Delta)$ and $\ln \frac{k_t}{\text{GeV}}$ of the declustering history of the hardest branch, as described above. 
In particular, in order to construct the image, we choose 25 by 25 pixels for both coordinates. Only the pixels corresponding to the declustering history are turned on. This way, image pixels have mainly two values either 1 or 0 and we do not need to normalize the data before using it for the CNN.

\subsubsection{CNN Architecture}
\begin{center}
\begin{figure}
\begin{tikzpicture}[node distance = 2cm, auto]
    \node [cloud] (input) {Input  $25 \times 25$};
    \node [block, below of=input] (firstCL) {Conv2D $N_1$ $N_F \times N_F$};
    \node [block, below of=firstCL] (secondCL) {Conv2D $N_2$ $N_F \times N_F$};
    \node [block, below of=secondCL] (thirdCL) {Conv2D $N_3$ $N_F \times N_F$};
    \node [block, below of=thirdCL] (fourthCL) {Conv2D $N_4$ $N_F \times N_F$};
    \node [process, below of=fourthCL] (linearL) {Dense Layer $N_n$};
    \node [decision, below of=linearL] (decide) {Predictions};
    \path [line] (firstCL) -- (secondCL);
    \draw [line] (secondCL) -- node[anchor=east] {Pooling} (thirdCL);
    \draw [line] (secondCL) -- node[anchor=west] {Dropouts} (thirdCL);
    \path [line] (thirdCL) -- node[anchor=west] {Dropouts}(fourthCL);
    \path [line] (fourthCL) -- node[anchor=east] {Pooling} (linearL);
    \path [line] (fourthCL) -- node[anchor=west] {Dropouts} (linearL);
     \path [line] (linearL) -- (decide);
    \path [line,dashed] (input) -- (firstCL);
\end{tikzpicture}
\caption{This is a schematic diagram of the CNN architecture used.  Here $N_i, i=1,..,4$ are the number of filters in each layer, $N_F$ is the filter size and $N_n$ is the number of neurons used in the flat layer.}\label{CNNcartoon}
\end{figure}
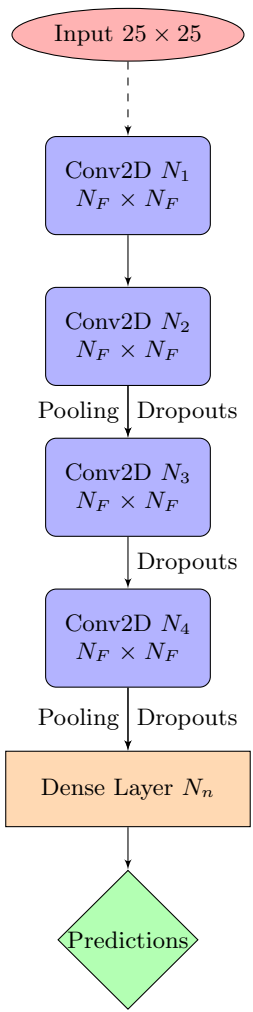
\end{center}

We employ convolutional NN for the signal-background classification, using the Lund plane images data set. \textsc{Keras}~\cite{keras} package is used for the CNN implementation. The CNN architecture is optimized for each benchmark. We use a balanced data set of 200K events for all the benchmarks. Each data set is divided into 60:20:20 proportions for training, validation and test set. In this section, we mention the generic set-up and details of the architecture which are common for all the benchmarks. A cartoon of the architecture is shown in Fig.~\ref{CNNcartoon}. We find that CNN with 4 convolutional layers of filter size 3 is the best choice for all cases. The number of filters for the convolutional layers is different in each case. After the convolutional layers, we have a flat layer with 800 neurons in all the cases. After the second and fourth convolutional layer, a pooling layer is used. For the robust training of the CNN, the down sampling of the feature maps is achieved by the pooling layers.  For the pooling layer we used the Max Pooling function. Another hyper parameter in the training is dropouts\footnote{It is a type of regularization to avoid over-fitting by dropping some connections in the network.}. We also turned on the dropout option with different strengths after the second, third and fourth convolutional layers. Further non-linearity is introduced in the model by using activation function `relu'. We use cross-entropy loss function. CNN training model parameters are updated using Adam optimizer~\cite{Kingma:2014vow} to minimize the loss function. Since for each data-set CNN architecture is optimised separately, further details about batch sizes and epochs and other information regarding the CNN will be mentioned later, in the respective sections.

\section{Analysis H $\to b \bar b$\label{hbb}}
In this section we discuss the $H \to b \bar b$ analysis. 
As previously mentioned, we are going to consider two scenarios: moderate-boost and high-boost. We will refer to the lower transverse momentum case as {\it benchmark point 1} (BP1), while the high-transverse momentum one will be BP2.

\begin{figure*}
\centering
\includegraphics[scale=0.55]{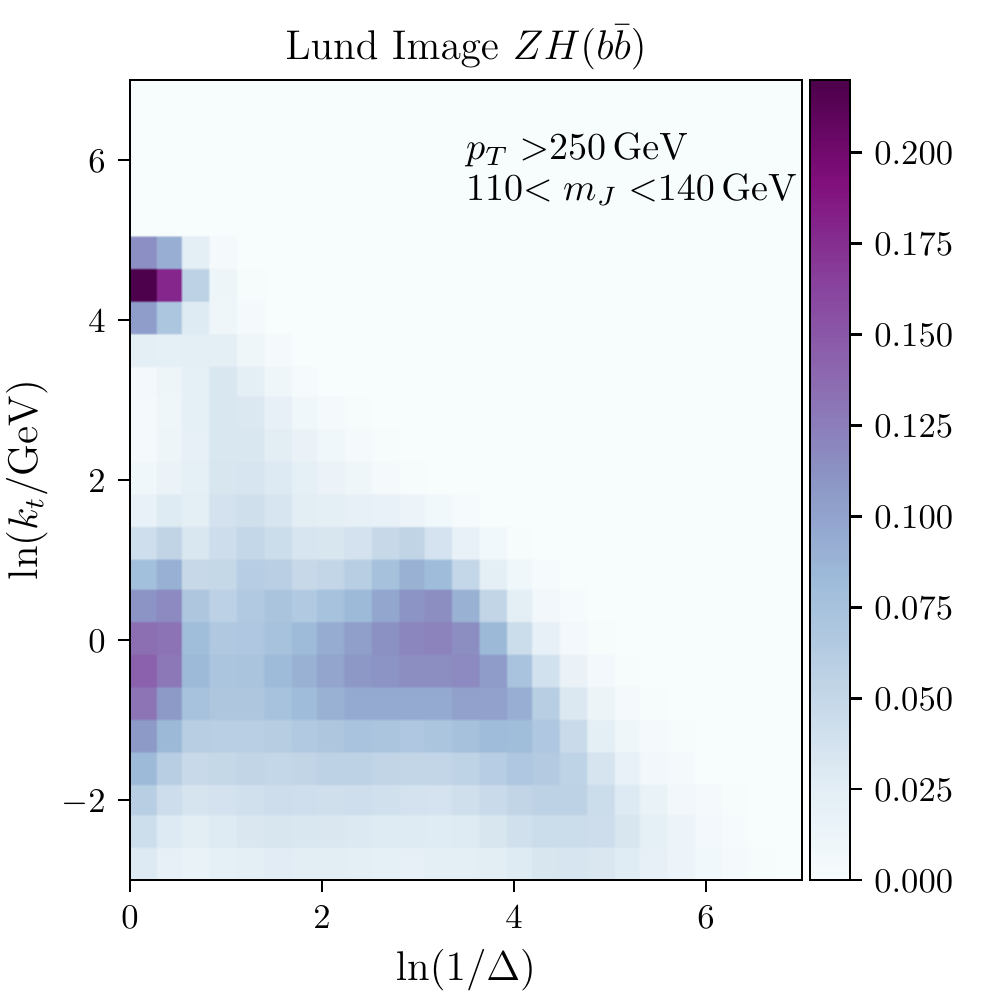}
\includegraphics[scale=0.55]{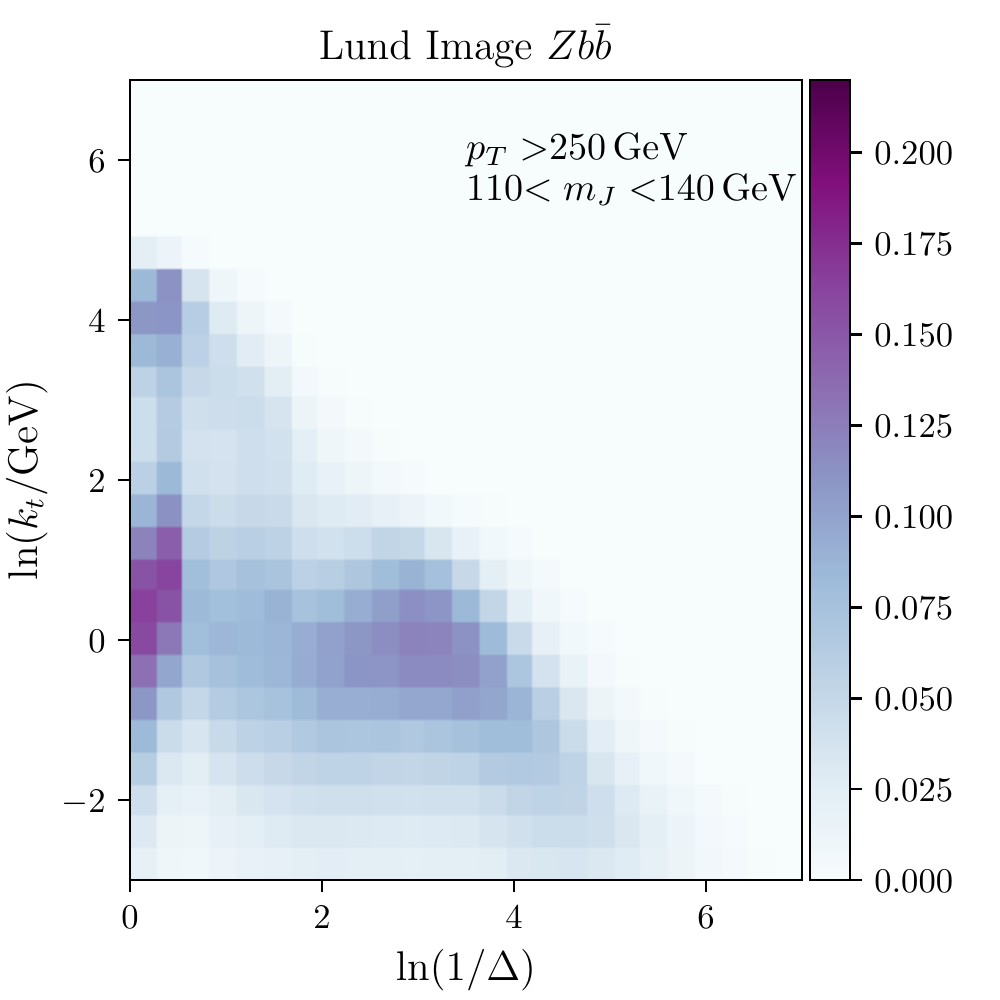}
\includegraphics[scale=0.55]{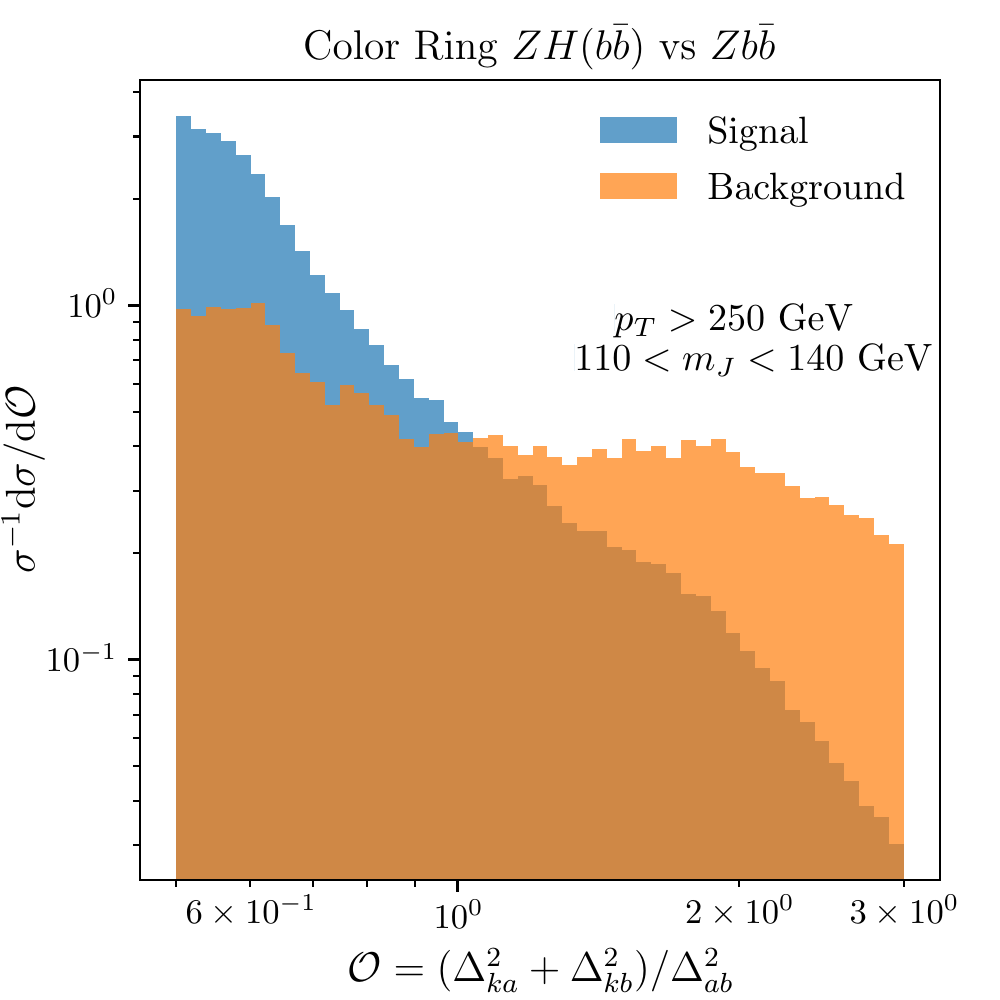}
\includegraphics[scale=0.55]{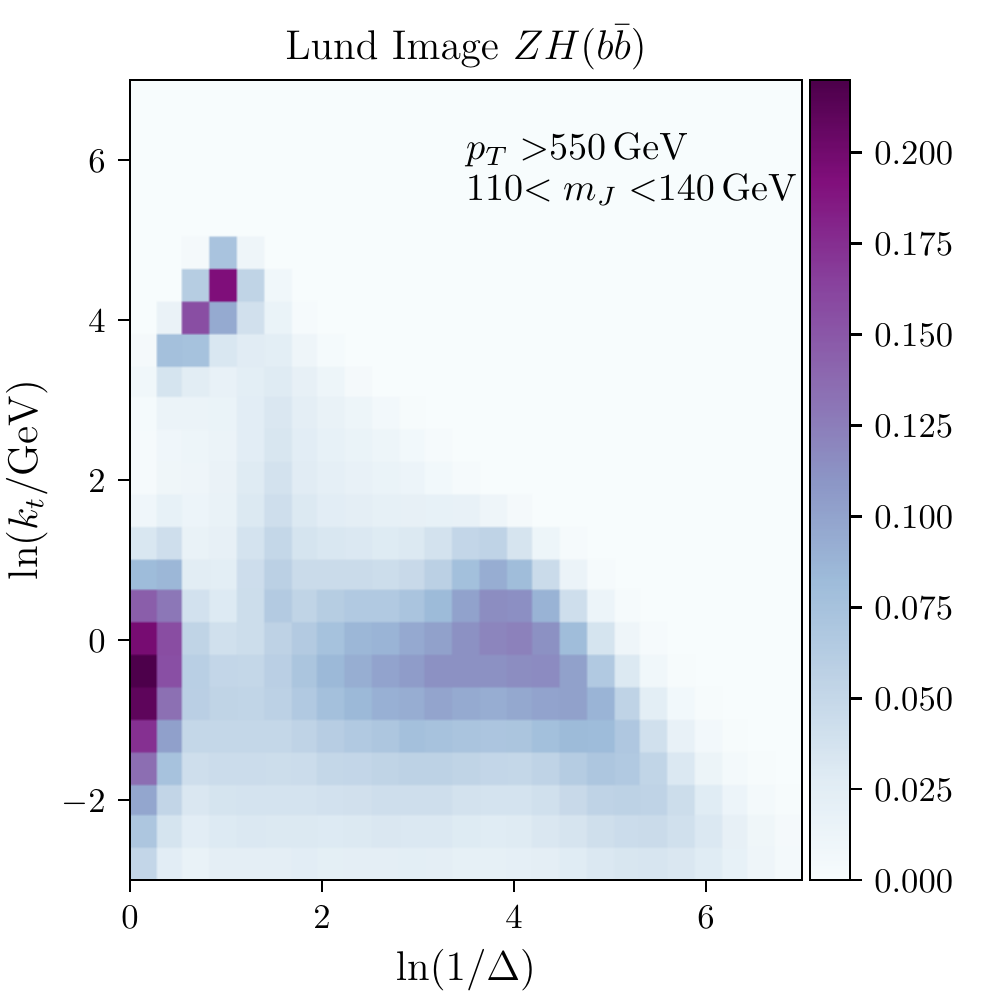}
\includegraphics[scale=0.55]{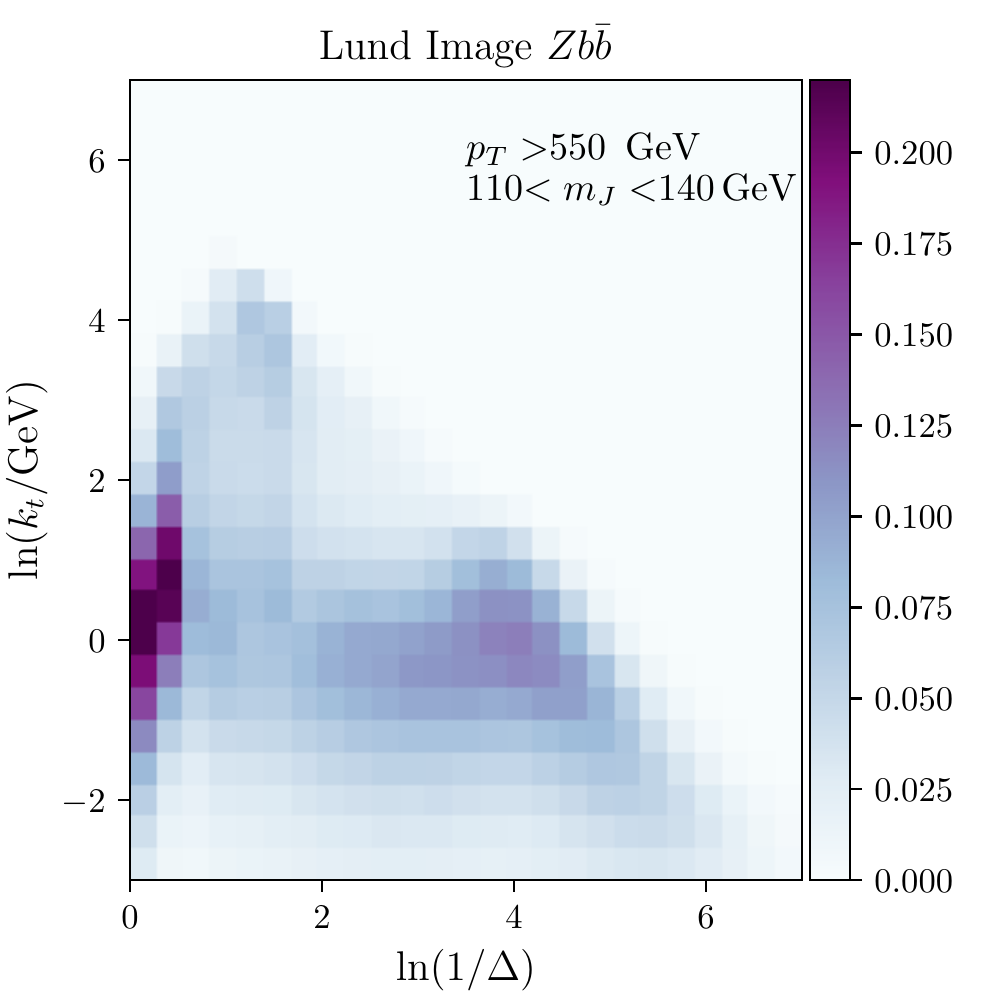}
\includegraphics[scale=0.55]{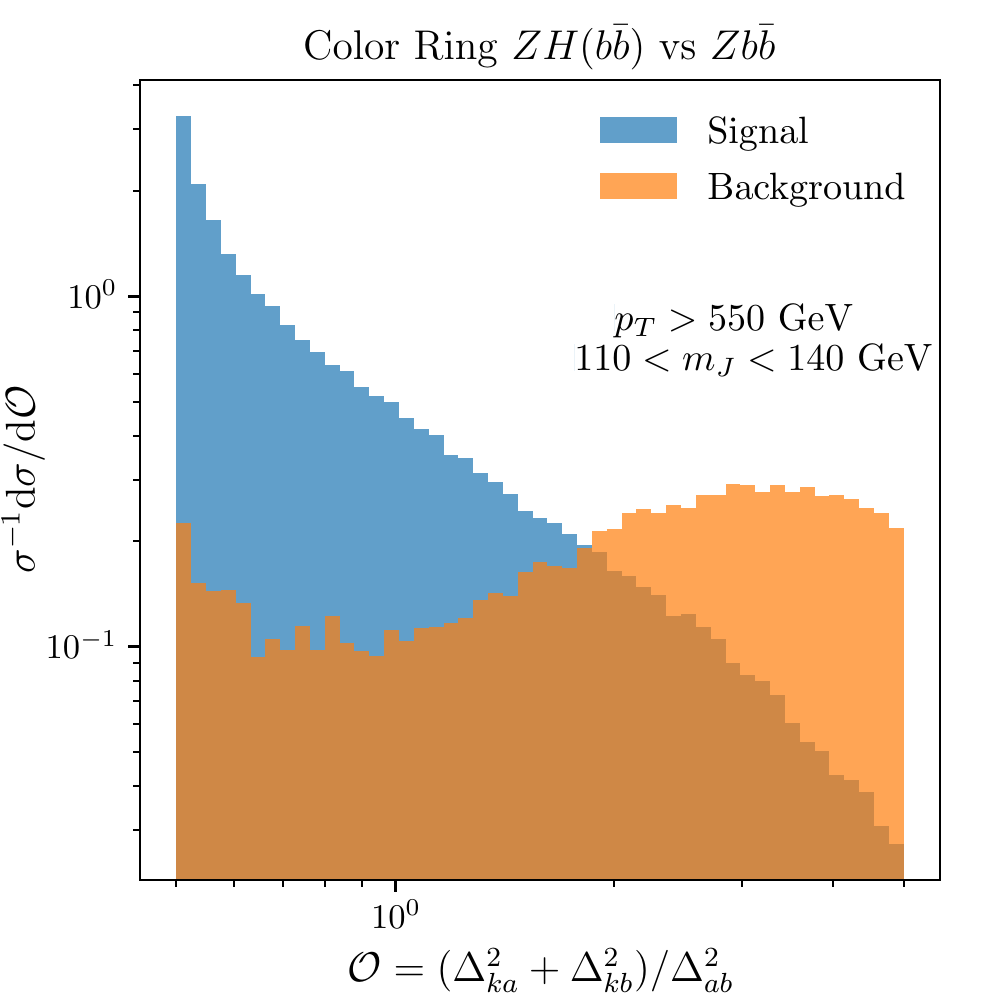}
\caption{The first and second columns show primary Lund jet plane images (averaged over 100K events) for $H \to b \bar b$ and $g \to b \bar b$ events, with jet $p_T>$ 250 GeV (upper panel) and $p_T>$ 550 GeV (lower panel). The image resolution is 25 $\times$ 25 for these images. The third column shows the normalized jet color ring distributions for the $H \to b \bar b$ and $g \to b \bar b$ events with jet $p_T>$ 250 GeV and $p_T>$ 550 GeV. In all cases, only jets with invariant mass $m_J \in [110,140] $ GeV are considered.}\label{hbbjmc_input}
\end{figure*}

\begin{figure*}
\centering
\includegraphics[scale=0.55]{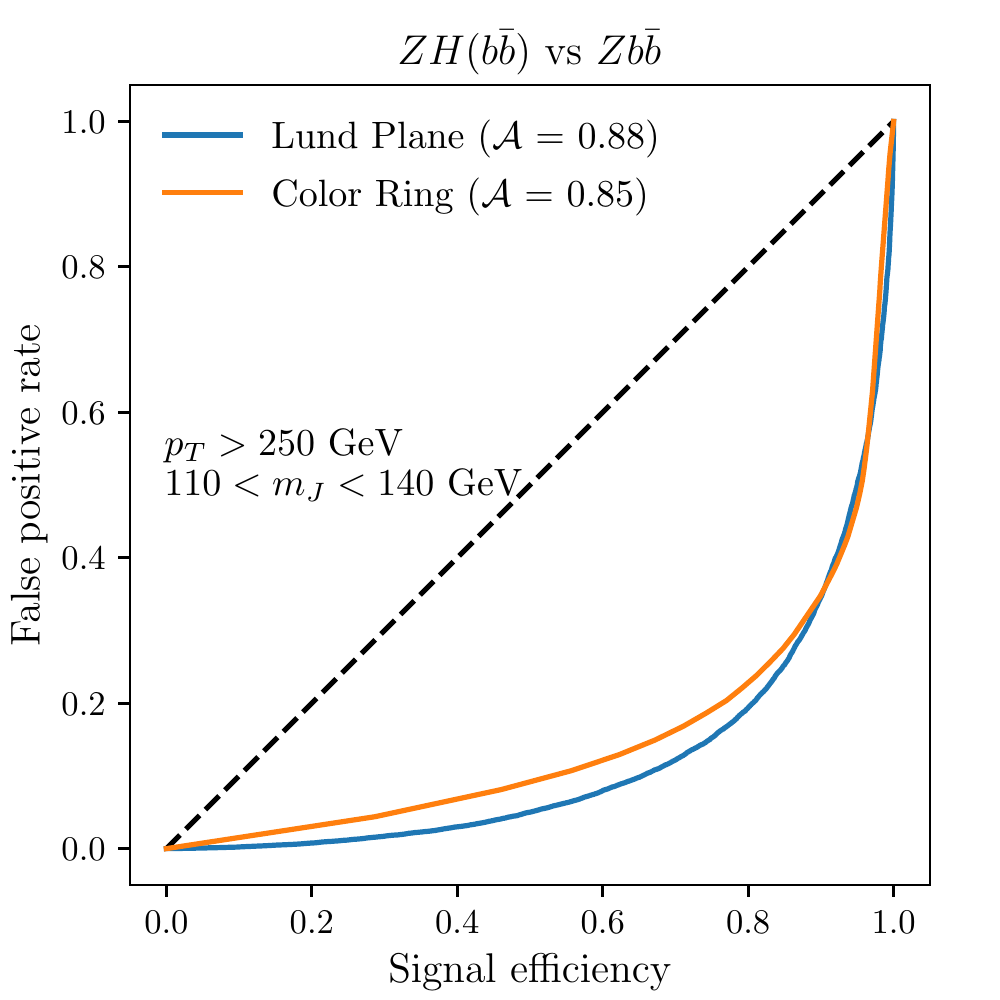}
\includegraphics[scale=0.55]{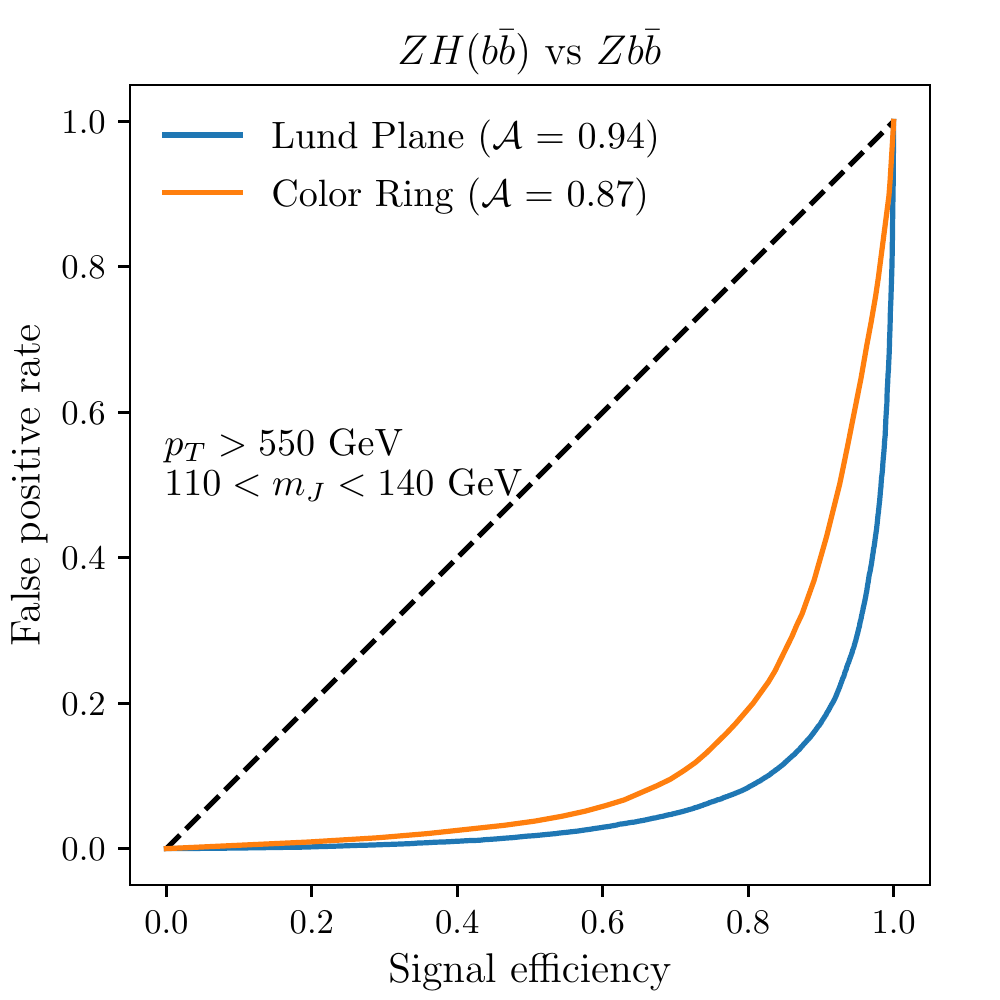}
\caption{ROC curves for CNN using the Lund jet plane images and for color ring for both the benchmarks of the $H \to b \bar b$ analysis considering jets with invariant mass $m_J \in [110,140] $ GeV.}\label{hbbjmc_ROCs}
\end{figure*}

\subsection{Moderate-boost scenario}
First, we analyse the moderate-boost scenario, where we require the transverse-momentum of the selected jet to be $p_T>$250 GeV.  As discussed in section~\ref{set-up}, we construct the primary Lund jet plane and the color ring observable, event by event. Averaged primary Lund jet plane images are then obtained considering 100K events. The resulting images are shown in the first row of Fig.~\ref{hbbjmc_input}, for signal jets (on the left) and background jets (in the center). The dominant differences between the Higgs image and the background one is clearly visible for large $\Delta$ and high $k_t$ $(\ln \frac{k_t}{\text{GeV}} \approx$4.5). Jet color ring distributions are instead shown in the third column of Fig.~\ref{hbbjmc_input}. Both signal and background distributions are normalized to unity. We note that signal events  mainly populate the  $\mathcal{O}<1$ region, while the background distribution is flatter, as expected~\footnote{Up to small differences in the simulation set-up, our color ring distributions are in good agreement with the ones obtained in Ref.~\cite{Buckley:2020kdp}.}.

We use CNN for the Lund jet images data set to perform the binary signal-background classification. The optimized CNN architecture has 4 convolutional layers, with filter size 3 and one flat layer with 800 neurons. The number of filters used is 16, for the first two convolutional layers, and 32 for the third and fourth layer. We use a batch size of 1000 and 15 epochs, i.e. \ the number of times the total data set is shown to the network, for CNN training. We did not need to train for a larger number of epochs because CNN trains faster than other deep learning methods. See Table~\ref{tab:CNNarchi} for more details of architecture (BP1). The Receiver Operating Characteristic (ROC) curves for CNN classification and for the color ring are shown in Fig.~\ref{hbbjmc_ROCs}, on the left. For the color ring case, we vary the threshold in small steps, considering the signal (background) below (above) the threshold, and calculate the signal efficiency and false positive rate for each value of the threshold to get the ROC curve. 

A standard metric used to assess the classification performance is the Area Under the ROC Curve (AUC). With our definition of ROC curves, optimal performance corresponds to AUC=0. In the following we will use the metric $\mathcal{A}=1-\text{AUC}$, where now $\mathcal{A}=1$ corresponds to optimal performance. 
We find that the performance of the CNN using the Lund jet plane data set is 3\% better than the single-variable approach using the color ring observable.
Assuming that the optimization of the CNN has been performed appropriately, the rather small improvement can be explained by the already good performance of the jet color ring, which was originally designed as an optimal color-singlet tagger.

\subsection{High-boost scenario}
We consider the second, boosted, scenario, with the leading jet transverse momentum required to be $p_T>$550 GeV. With the exception of the generation-level $p_T^{\mu\mu}$ and leading-jet $p_T$ cut in the analysis, all the set-up is the same as in the previous case. The overall analysis selection efficiency for the background events is higher in this case as compared to the previous case. The averaged (over 100K events) primary Lund jet plane images for the signal and background, with jet mass cut $110<m_J<140$~GeV, are shown in the second row of Fig.~\ref{hbbjmc_input}. 

The most prominent difference in the signal image, with respect to the moderate-boost scenario, is  
the shift of the high-$k_T$ patch towards smaller values of $\Delta$, and hence larger values of $-\ln \Delta$. This happens because the decay products of the Higgs boson tend to be more collimated.
The background image also noticeably changes with respect to the lower-$p_T$ case.
The color ring signal and background distributions are shown in the lower right column in Fig.~\ref{hbbjmc_input}.


The CNN architecture details are mentioned in the Table~\ref{tab:CNNarchi} under BP2 column. This architecture has a lesser number of filters in the third and fourth convolutional layer than compared to the moderate-boost scenario. In the right panel of Fig.~\ref{hbbjmc_ROCs}, we show the ROC curves for the CNN predictions for the Lund images and for the color ring observable. In this case, the classification accuracy of the Lund plane+CNN combination is significantly better (7\%) than the single-variable approach using color ring observable. The color ring $\mathcal{A}$ is 0.02 more than the one of BP1, but the CNN classification accuracy is 6\% better than the moderate boost case.

\begin{figure*}
\centering
\includegraphics[scale=0.55]{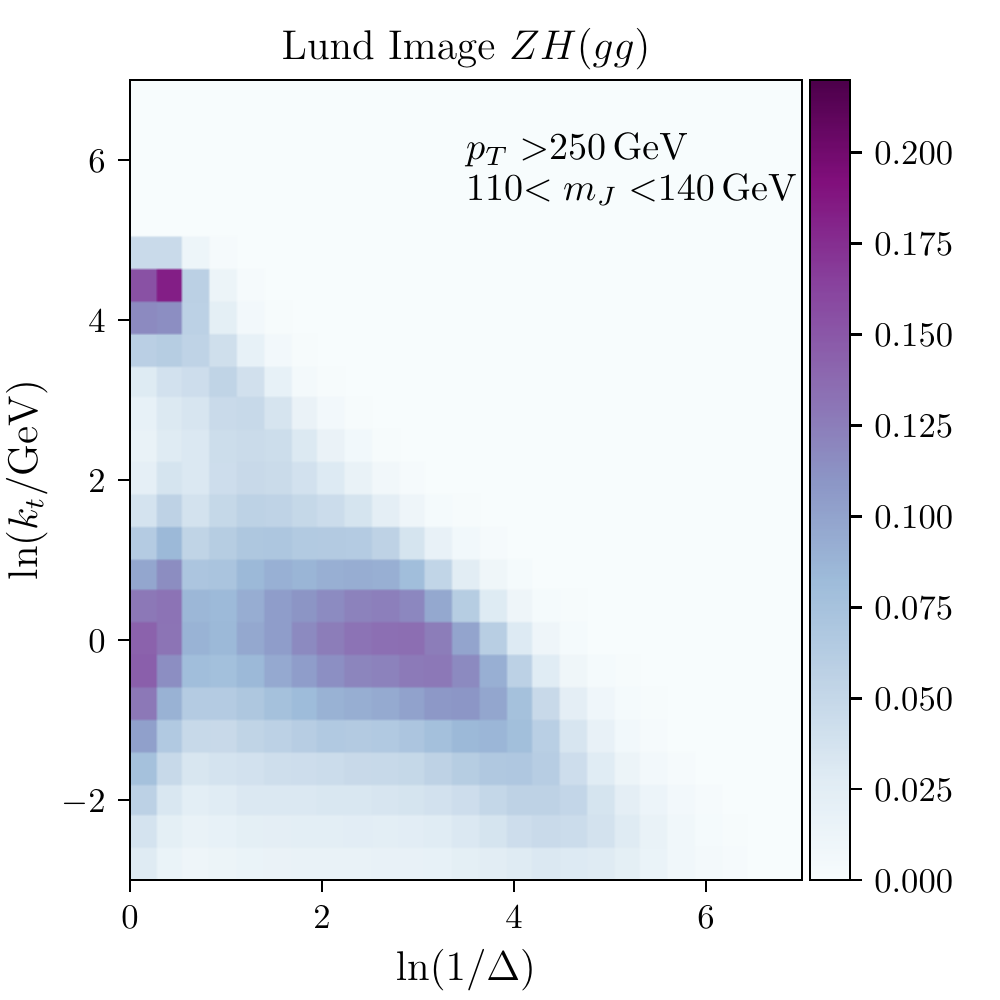}
\includegraphics[scale=0.55]{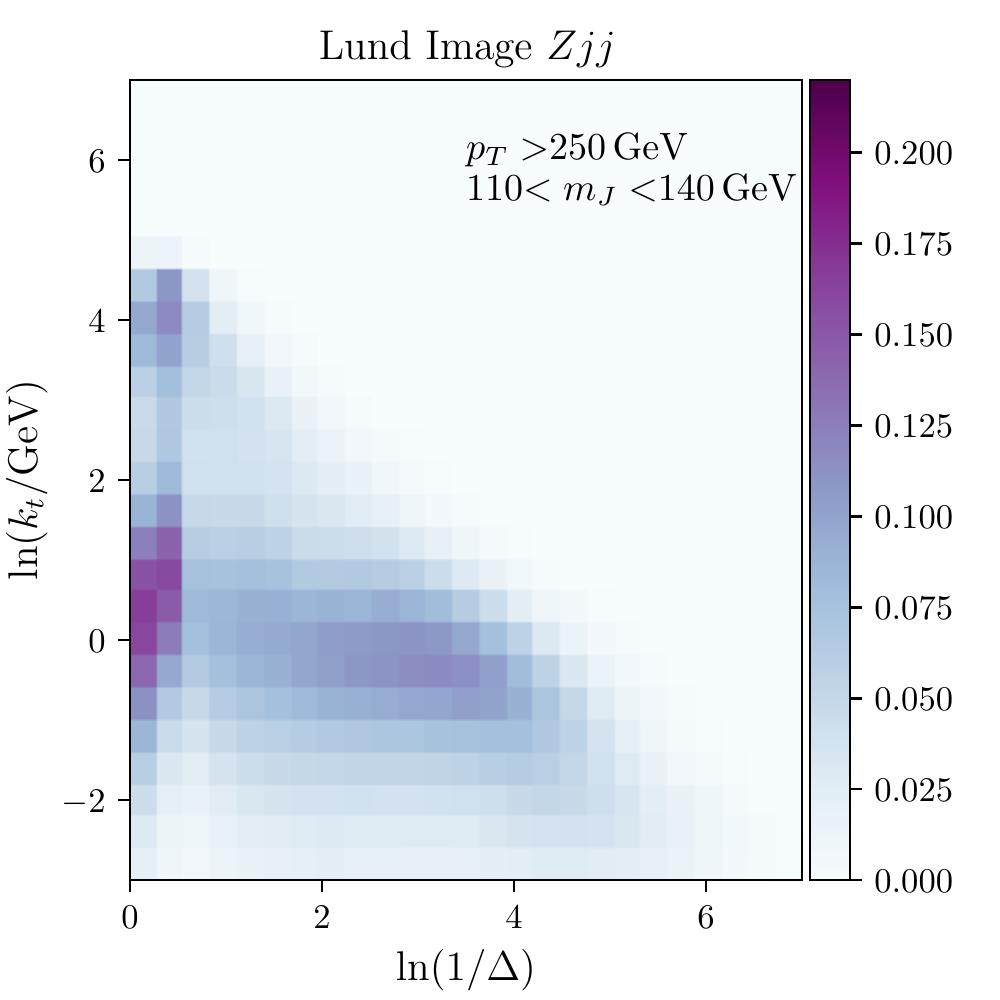}
\includegraphics[scale=0.55]{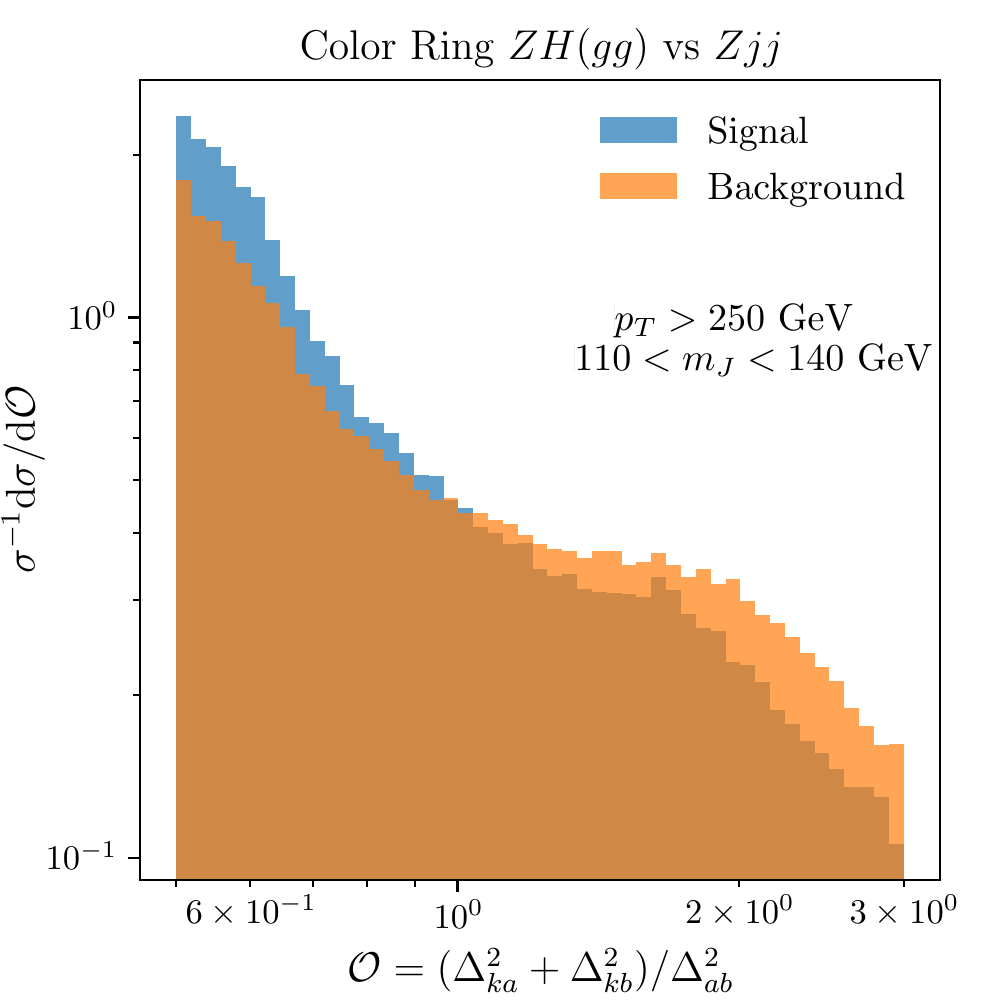}
\includegraphics[scale=0.55]{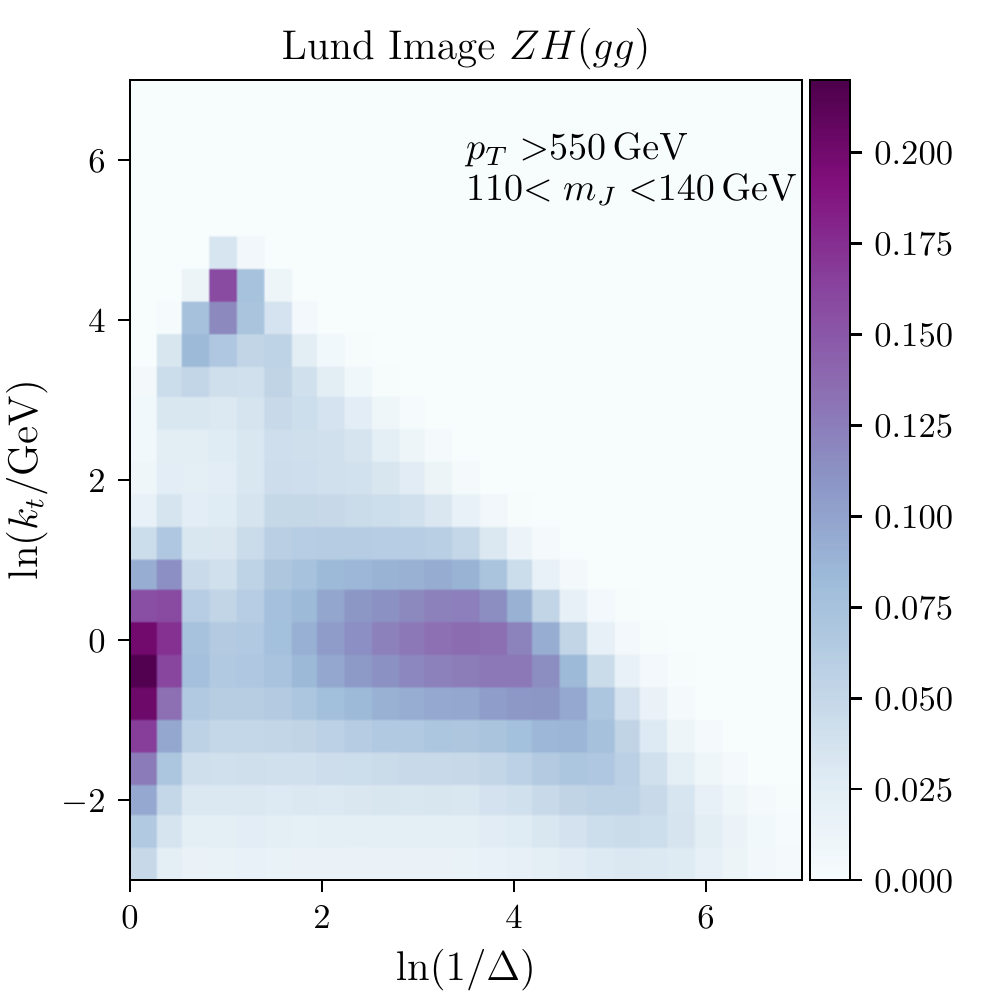}
\includegraphics[scale=0.55]{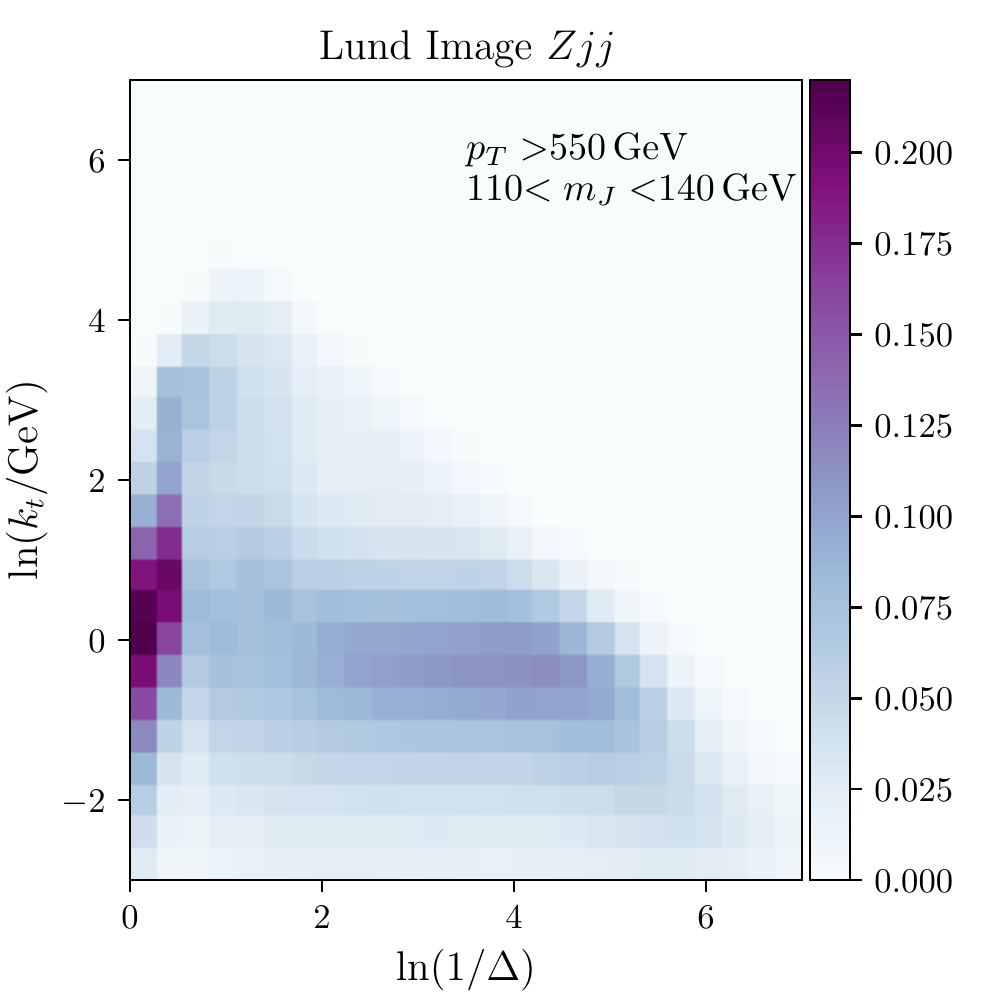}
\includegraphics[scale=0.55]{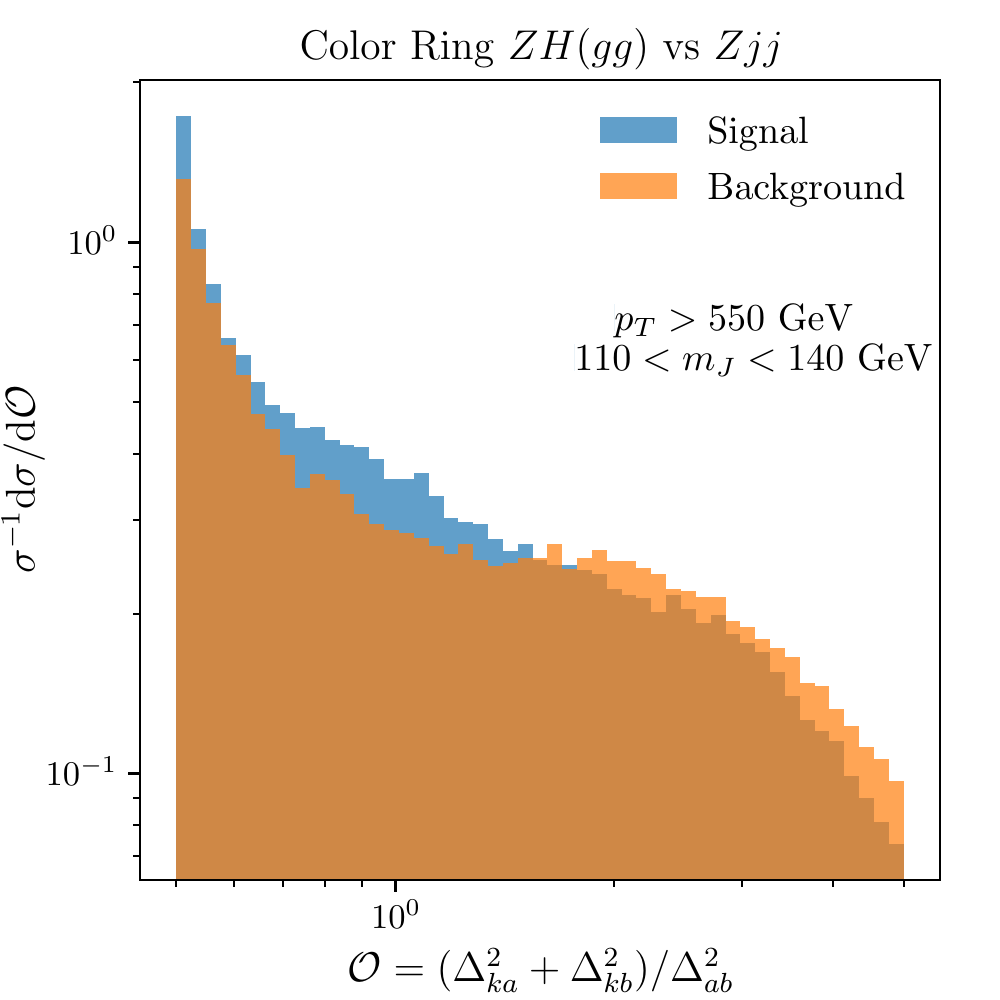}
\caption{The first and second columns show primary Lund jet plane images (averaged over 100K events) for $H \to gg$ and $Zjj$ events, with jet $p_T>250$~GeV (upper panel) and $p_T>550$~GeV (lower panel). The image resolution is 25 $\times$ 25 for these images. The third column shows the normalized jet color ring distributions for the $H \to gg$ and $Zjj$ events with jet $p_T>250$~GeV and $p_T>550$~GeV. In all cases, only jets with invariant mass $m_J \in [110,140] $ GeV are considered.}\label{hggjmc_input}
\end{figure*}

\begin{figure*}
\centering
\includegraphics[scale=0.55]{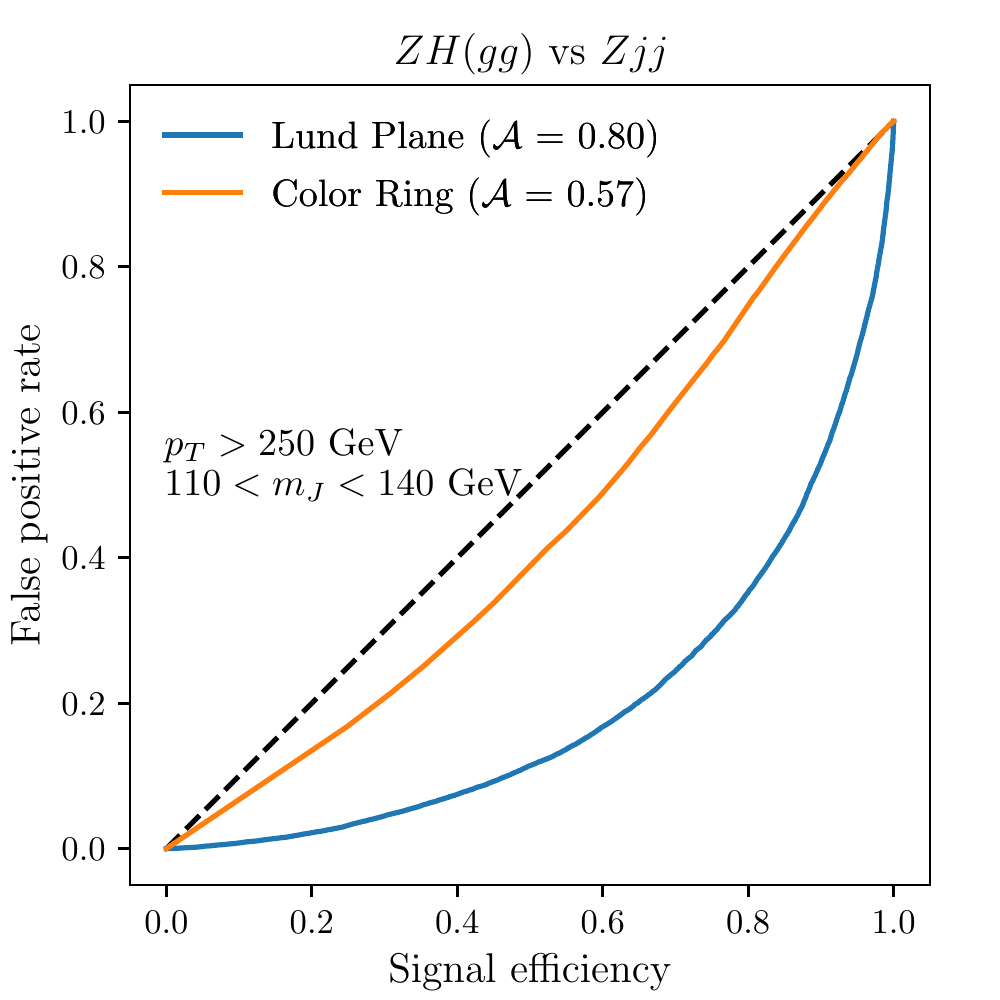}
\includegraphics[scale=0.55]{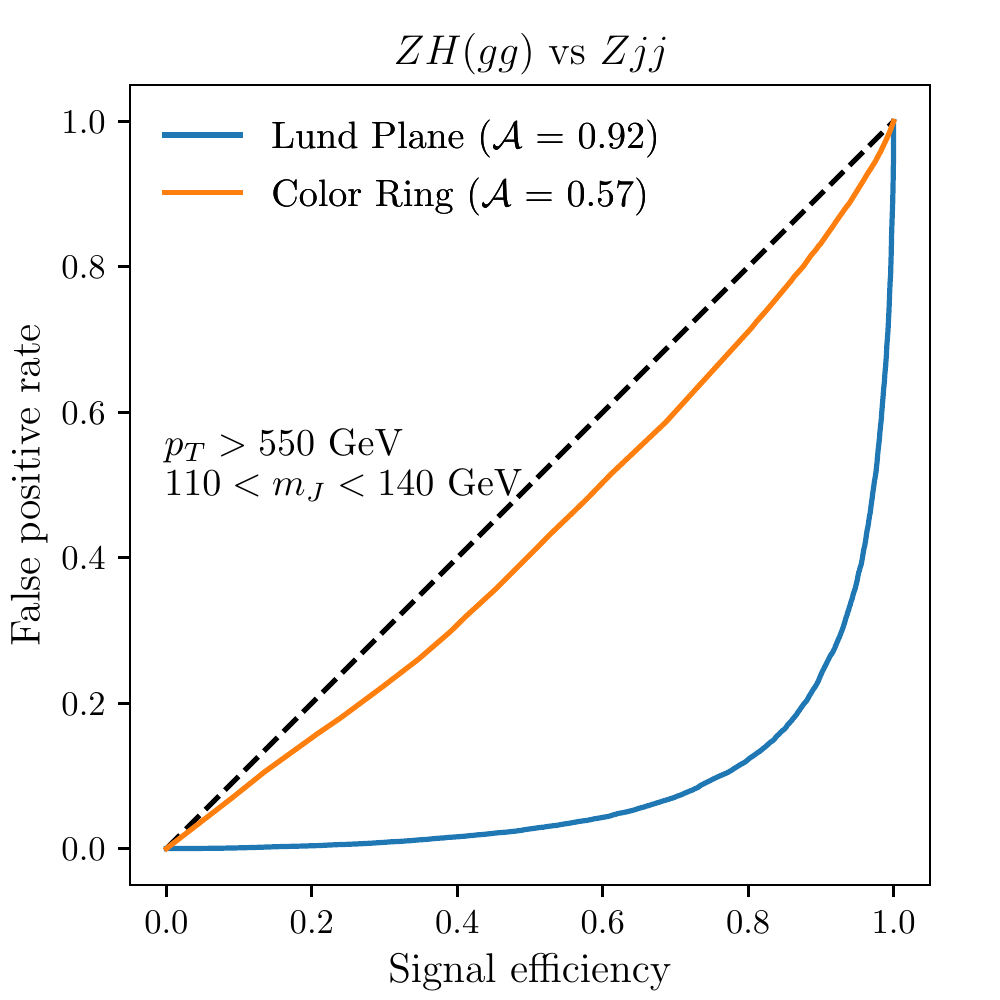}
\caption{ROC curves for CNN using the Lund jet plane images and for color ring for both the benchmarks of the $H \to gg$ analysis considering jets with invariant mass $m_J \in [110,140] $ GeV.}\label{hggjmc_ROCs}
\end{figure*}

\section{Analysis H $\to gg$\label{hgg}}
We now move to analyse the other decay channel of the Higgs boson considered in this study, namely the light-jet final-state. This is a very challenging decay channel and, as mentioned earlier, the jet color ring is known to perform poorly in this context.

We use the same tools to simulate the events for $H \to gg$ benchmark as in the previous case except the model used in \textsc{Madgraph}. This decay mode of Higgs is mediated by the heavy quark loop and $H \to gg$ effective coupling is implemented in HEFT model~\cite{heft}. For consistency, we use the same model to simulate the background events i.e. $pp \rightarrow$ $Z(\mu^+\mu^-)+jj$. In this section, we follow the same ordering as $H \to b \bar b$ analysis i.e first $p_T>250$ GeV (BP3),  and followed by $p_T>550$ GeV (BP4).

\subsection{Moderate-boost scenario}
Using the analysis set-up of section~\ref{set-up}, we form a primary Lund jet plane and color ring for signal and background events. The averaged Lund jet plane images and color ring distributions are shown in the first row of Fig.~\ref{hggjmc_input} for the $p_T>250$ GeV benchmarks. We start by considering the jet color ring. Similarly, to the $H \to b \bar b$ analysis, the signal distribution is falling sharply, i.e.\ signal events mainly populate in $\mathcal{O}<1$ region, in accordance to our expectation for color-singlets. However, as it was found in Ref.~\cite{Buckley:2020kdp}, the behavior of the background distribution is very different from the corresponding $H \to b\bar b$ case and it is in fact almost overlapping with the signal distribution. In this case, several possible color configurations are contributing, while the previous $g \to b \bar b$ case was characterized by the octet configuration. 

It is then interesting to check whether the Lund jet plane can instead highlight differences between signal and background in $H\to gg$. As we can see from the first two plots in Fig.~\ref{hggjmc_input} this is indeed the case. We can make a more quantitative statement about the tagger performance by looking at the  ROC curves for the Lund jet plane + CNN and color ring. This is done in Fig.~\ref{hggjmc_ROCs}, on the left. Then CNN using Lund jet plane data set provides very good classification performance, while the color ring ROC is very poor, close to the random classifier.

\subsection{High-boost scenario}
A very similar picture holds for the high-boost scenario, $p_T>550$~GeV case. Corresponding results are reported in the second row of Fig.~\ref{hggjmc_input}. In particular,  the differences in the averaged Lund jet plane images are clearly visible by eye. For instance, as in the $H \to b \bar b$ case, there is a bright patch at high $k_t$. Lund jet plane and color ring ROC curves for this benchmark are shown in Fig.~\ref{hggjmc_ROCs}, on the right.

\begin{figure*}
\centering
\includegraphics[scale=0.6]{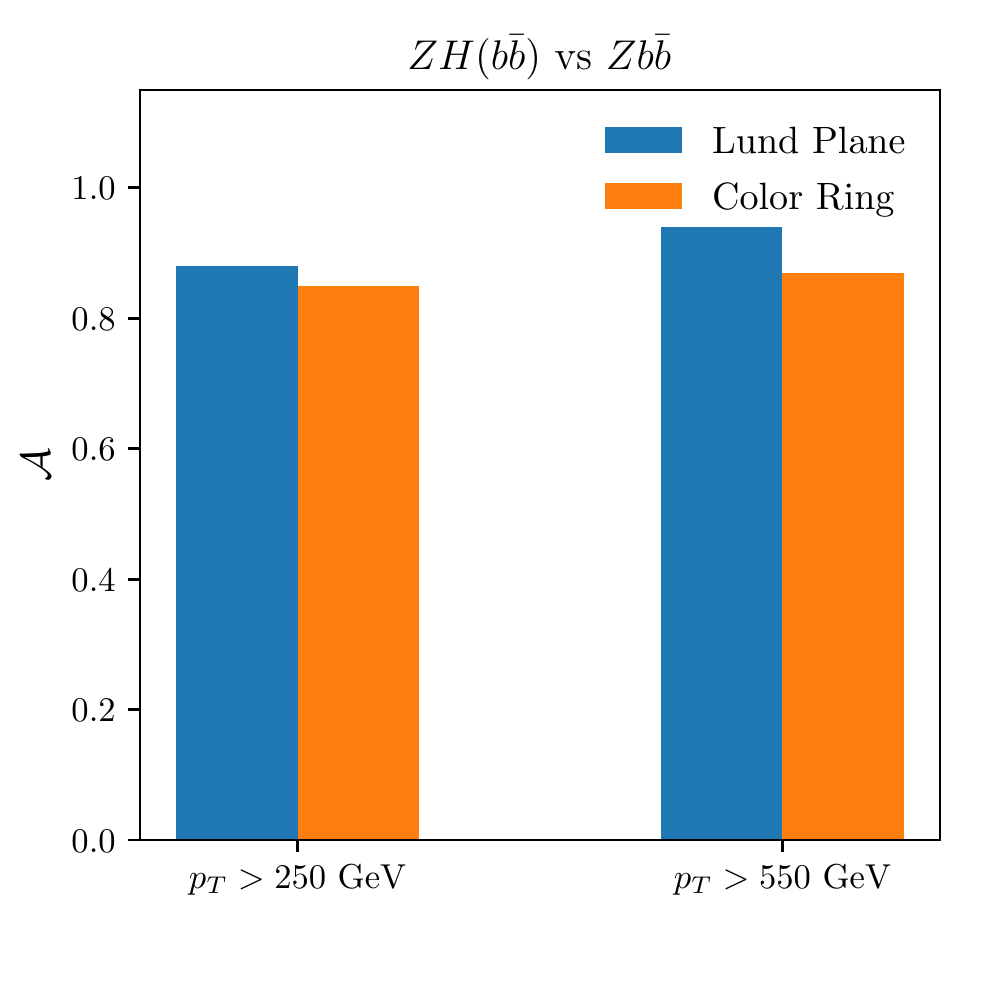}
\includegraphics[scale=0.6]{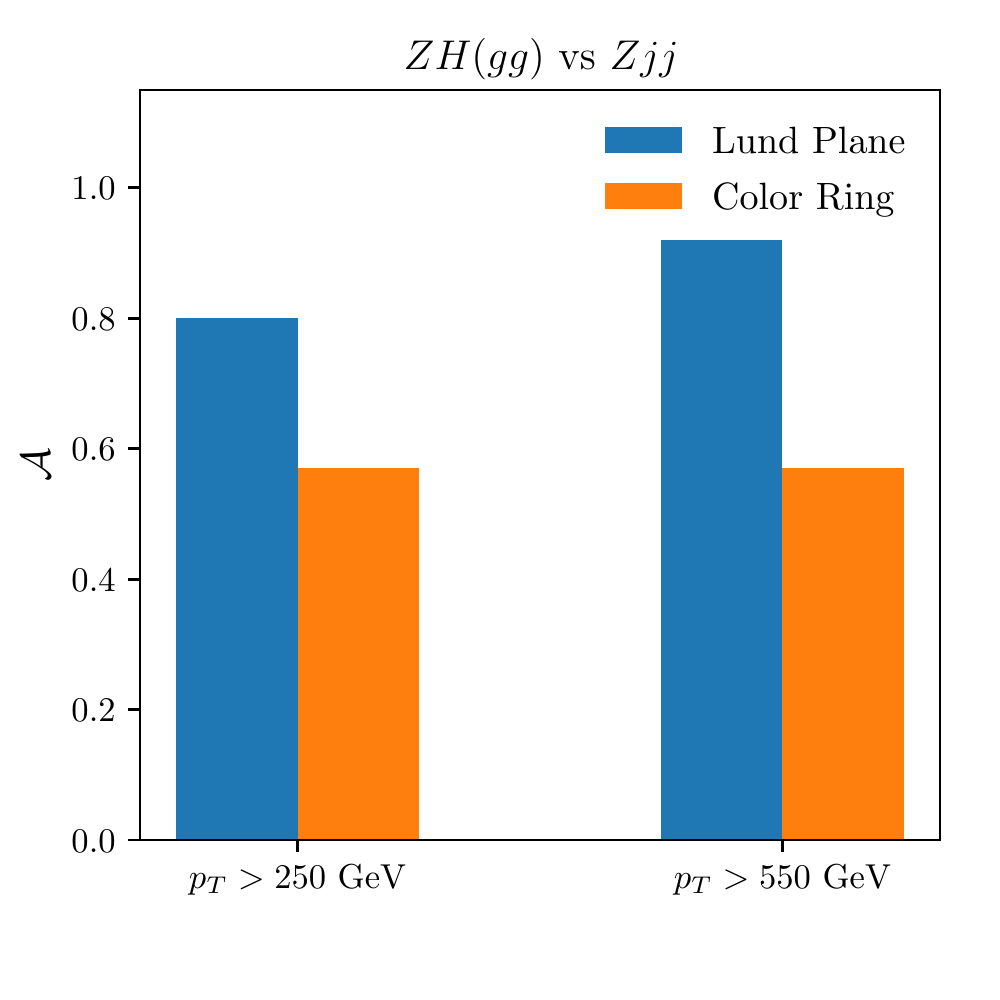}

\caption{Left: Higgs signal and background classification performance comparison using color ring and Lund plane+CNN combination for the $H \to b \bar b $ analysis. Right: same comparison for the $H \to gg$ analysis.}\label{performance}
\end{figure*}

\section{Conclusion and Outlook\label{conclusion}}
In this work, we have studied the primary Lund jet plane in the context of tagging hadronically decaying Higgs bosons, in the boosted regime, where the Higgs' decay products are reconstructed into a single large-radius jet.
In particular, we have considered two challenging, but crucial, decay channels: the heavy flavor (bottom) one and the Higgs decay into light jets. 
We have concentrated on two transverse momentum benchmark scenarios, namely moderate and high boost of the Higgs boson. 

 Inspired by previous work on $W$ and top tagging using the Lund jet plane~\cite{Dreyer:2018nbf,Dreyer:2020brq}, we have built images that are used as inputs to a convolutional neural network for classification.  
 We have compared the performance of this tagger to a more standard approach: a single-variable analysis that exploits a theoretical-motivated observable, namely the jet color ring. 
 
 Our findings, for different transverse momentum scenarios, are summarised in Fig.~\ref{performance}, for the $H \to b\bar b$ and $H \to gg$ analysis, respectively. 
 We have taken $\mathcal{A}=1-\text{AUC}$ as figure of merit to assess the taggers' performance. 
 We can see that the Lund jet plane and CNN combination has the best separation power for all the benchmarks studied. 
 Its performance is equally good for $H\to b \bar b$ and the $H \to gg$ analyses, thus providing us with some confidence about its robustness. This is in stark contrast with the jet color ring, which almost equals the CNN performance in the $H\to b \bar b$ case, while fails completely for the $H\to gg$ process. 
 We note that the difference between the AUC for CNNs and color ring cases is higher for the high-boost case, in both analyses.

Despite the fact that raw information from the jet (or even particles') kinematics can serve as inputs to ML algorithms, the use of the Lund jet plane is intriguing from a theoretical perspective. It provides a physically-motivated picture of a jet that can be naturally used as input to a NN. At the same time, the Lund jet plane can be described with perturbative field theory. Indeed, the Lund plane density has been recently computed in QCD, for light jets~\cite{Lifson:2020gua}. It would be extremely interesting to perform analogous calculations for signal jets and $b$-jets, in order to shed light on the taggers' performance, as it was done, for instance, in the case of single-variable taggers almost a decade ago, e.g.~\cite{Dasgupta:2013ihk,Dasgupta:2013via,Dasgupta:2015yua}. 

Furthermore, one important feature of the Lund jet plane, especially when expressed in terms of the variable $k_t$, as in our case, is that it makes the separation between regions dominated by perturbative and non-perturbative physics rather clear. If we assume that the latter is characterized by energy scales corresponding to relative transverse momenta of the order of 1~GeV, then the boundary between the two regions is a straight, horizontal, line at $\ln (k_t/ \text{GeV})=0$. Even by eye, we note that, in all the cases considered in this study, the bulk of the difference between signal and background Lund jet plane images is found above this horizontal line. Thus, we expect that the CNN is mostly exploiting perturbative information. We have confirmed this intuition by performing an analysis where we input to the CNN only the upper section of the Lund jet plane, i.e.\ $\ln (k_t/ \text{GeV})>0$. We have found the difference between the values of $\mathcal{A}$ in the two cases to be below the percent level.~\footnote{We thank Andrew Larkoski for suggesting this study.}

Finally, in this work we used simulated data without incorporating detector effects. In a realistic situation, we expect some degradation in the reported results.  
It would be then important to study the resilience~\cite{Bendavid:2018nar} of the color ring and of the Lund jet plane against this type of contributions. 
On the other hand, one could improve the performance of the taggers by exploiting more information from the clustering history.
This can be achieved, for instance, by going beyond the primary plane approximation.
Furthermore, the use of other ML architectures, such as, e.g.\ graph neural networks, may lead to a further gain in the classification performance~\cite{Dreyer:2020brq}.
We plan to explore these directions in the future work.

\section{acknowledgments}
We thank Andrea Coccaro, Fr\'ed\'eric Dreyer, Andrew Larkoski, and Giovanni Stagnitto for comments on the manuscript.
We also thank the members of the ATLAS groups in Genova and Pavia for many useful discussions on Higgs tagging and color flow. 
This work is supported by Universit\`a di Genova under the curiosity-driven grant ``Using jets to challenge the Standard Model of particle physics'' and by the Italian Ministry of Research (MUR) under grant PRIN 20172LNEEZ.
\appendix 
\newcommand{\hbAppendixPrefix}{A}
\renewcommand{\thefigure}{\hbAppendixPrefix\arabic{figure}}
\setcounter{figure}{0}

\setcounter{table}{0}
\section{CNN architectures\label{cnn2}}
\begin{figure*}
\centering
\includegraphics[width=0.45\textwidth]{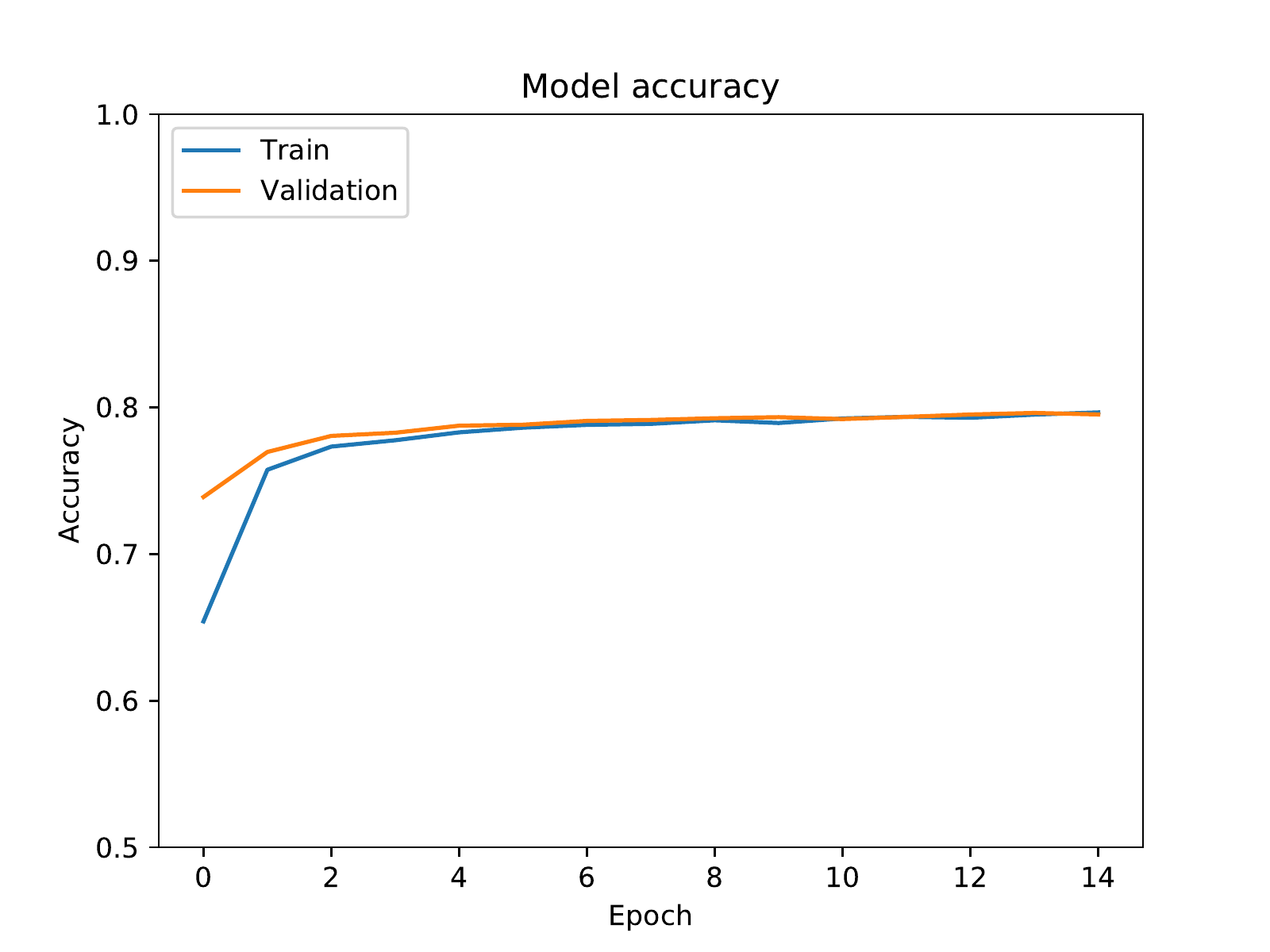}
\includegraphics[width=0.45\textwidth]{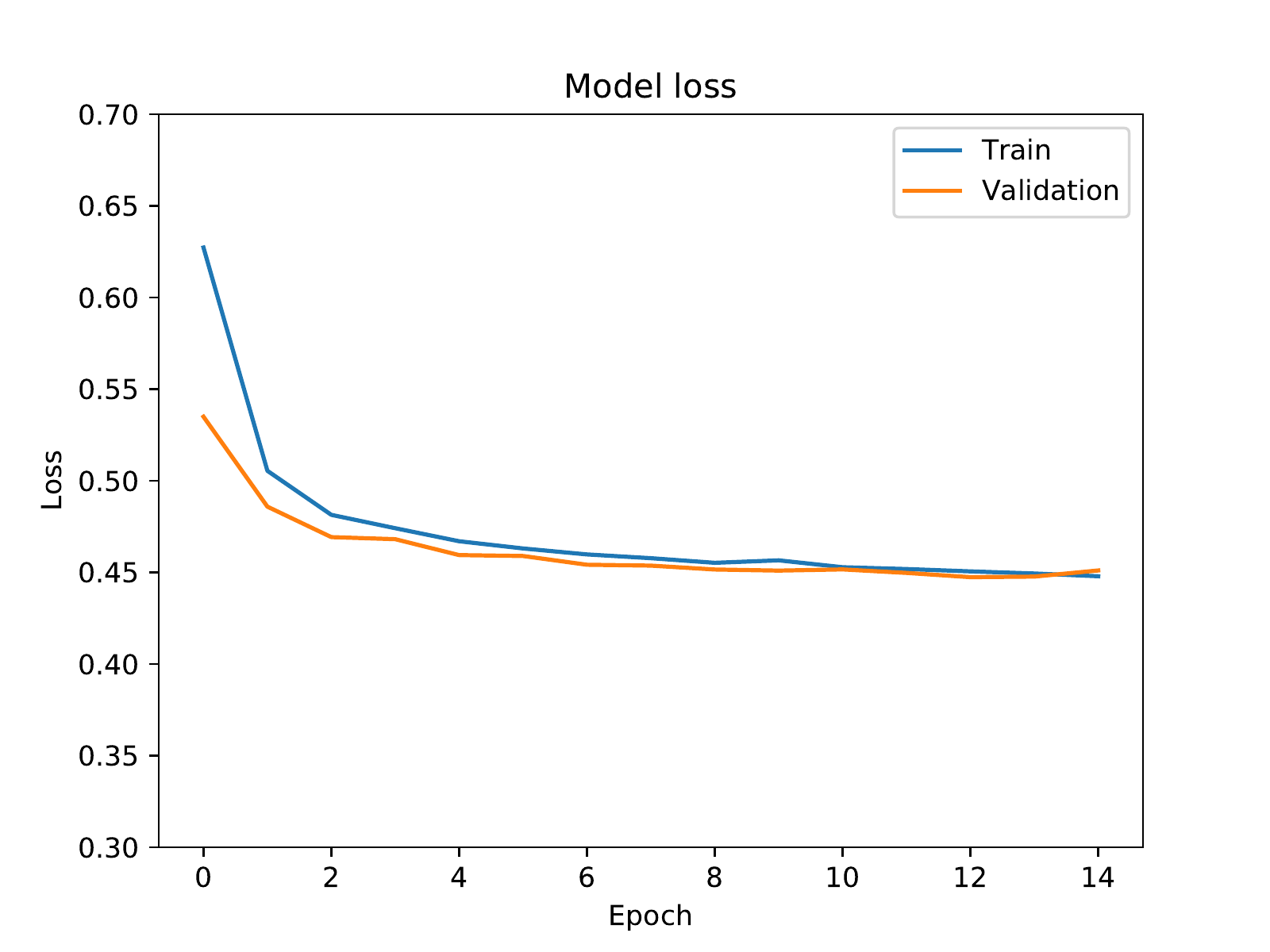}
\includegraphics[width=0.45\textwidth]{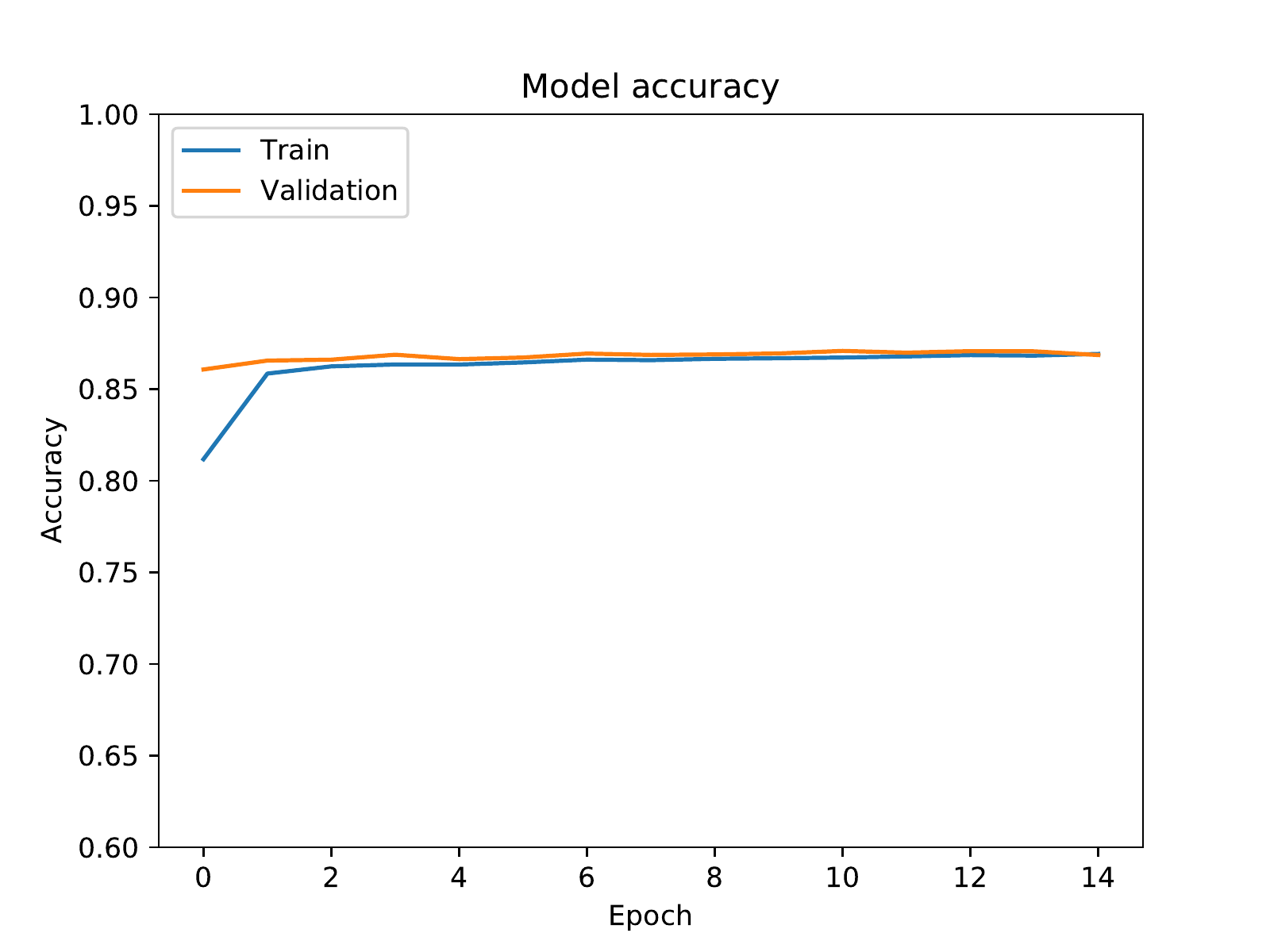}
\includegraphics[width=0.45\textwidth]{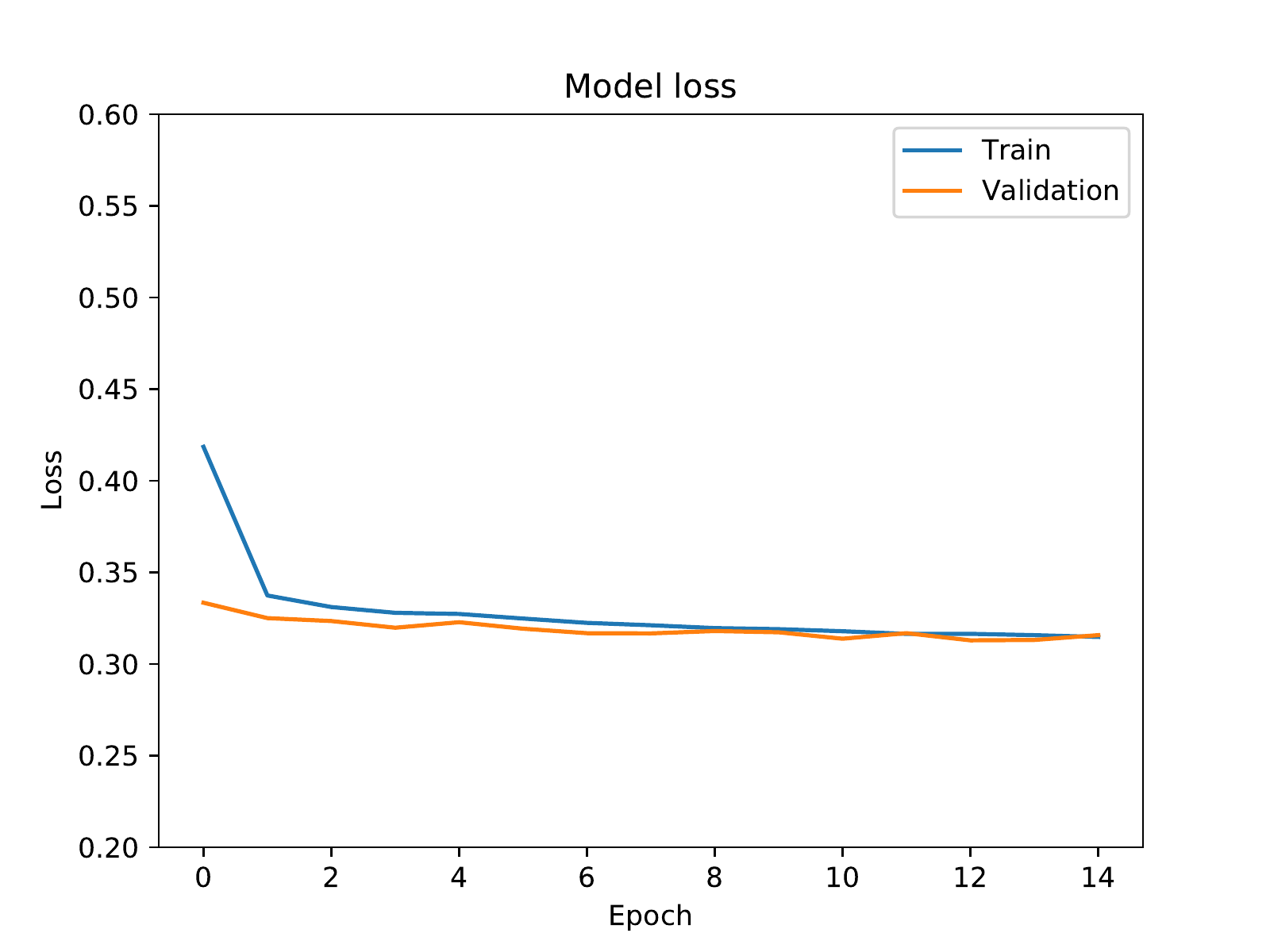}
\caption{\label{fig:training1}CNN training curves for BP1, on the top, and for BP2, at the bottom.}
\end{figure*}

The main structure of the architecture (see Fig.~\ref{CNNcartoon}) is the same for all cases. For each benchmark, we ran several models with different choices of hyper-parameters and presented the result for the best ones, which we call optimised models.
As an example, model training curves for BP1 and BP2 are shown in Fig.~\ref{fig:training1}. 
Since all benchmark datasets share some common features, the resulting optimised architectures are also quite similar. 
We optimise the architecture for each benchmark because our purpose is to see the maximum performance gain over the single variable case. 
Thus, even if in an actual experimental analysis, one  decides to use one model for all the data (which means for all the benchmarks considered here), we do not expect significant reduction in the classification performance. In Table \ref{tab:CNNarchi}, we include all the details about the architecture for each benchmark. The first two benchmarks are for the $H \to b \bar b$ analysis and last two are for the $H \to gg$ analysis. They are ordered by $p_T$ of the leading jet. 
$N_1$, $N_2$, $N_3$, and $N_4$ are the number of filters in the first, second, third and fourth convolutional layer, respectively. A filter size of $3 \times 3$ is used in all the cases. Further filter movement to analyze the image is controlled by the stride parameter, which we choose as one unit along both directions. In all the cases, we zero padded the images at third and fourth layer in such a way that the input and output image dimensionality remains the same.%

\begin{table}
\centering
\begin{tabular}{lcccccc}
\hline
\hline & BP1 & BP2  & BP3 & BP4 \\
\hline
\hline
$N_1$ Conv2D   & 16    &  16        &  16   &  16      \\
$N_2$ Conv2D   & 16    &  16       &  16   &  16     \\
Dropouts       & 0.25  &  0.05    & 0.20   &  0.20    \\
$N_3$  Conv2D  &  32   &  16        &  16   &  32      \\
$N_4$  Conv2D  & 32    &  16  &  16   &  32      \\
Dropouts       &-      &  0.05   &  0.30  &  0.30      \\
Flat Layer   & 800   & 800       &  800   & 800    \\
Epochs         & 15    & 15      &  20   &  20   \\
Batch Size     & 1000  & 1000     &  800   &  700    \\
\hline
\hline
\end{tabular}
\caption{CNN architectures used for different data sets. }
\label{tab:CNNarchi}
\end{table}

\begin{thebibliography}{99}

\bibitem{Marzani:2019hun}
S.~Marzani, G.~Soyez and M.~Spannowsky,
Lect. Notes Phys. \textbf{958} (2019), pp,
[arXiv:1901.10342 [hep-ph]].


\bibitem{Kasieczka:2019dbj}
G.~Kasieczka, T.~Plehn, A.~Butter, K.~Cranmer, D.~Debnath, B.~M.~Dillon, M.~Fairbairn, D.~A.~Faroughy, W.~Fedorko and C.~Gay, \textit{et al.}
SciPost Phys. \textbf{7} (2019), 014,
[arXiv:1902.09914 [hep-ph]].

\bibitem{Freitas:2019hbk}
F.~F.~Freitas, C.~K.~Khosa and V.~Sanz,
Phys. Rev. D \textbf{100} (2019) no.3, 035040,
[arXiv:1902.05803 [hep-ph]].

\bibitem{Brehmer:2018kdj}
J.~Brehmer, K.~Cranmer, G.~Louppe and J.~Pavez,
Phys. Rev. Lett. \textbf{121} (2018) no.11, 111801,
[arXiv:1805.00013 [hep-ph]].

\bibitem{Brehmer:2018eca}
J.~Brehmer, K.~Cranmer, G.~Louppe and J.~Pavez,
Phys. Rev. D \textbf{98} (2018) no.5, 052004,
[arXiv:1805.00020 [hep-ph]].

\bibitem{Chen:2020mev}
S.~Chen, A.~Glioti, G.~Panico and A.~Wulzer,
[arXiv:2007.10356 [hep-ph]].

\bibitem{Kasieczka:2018lwf}
G.~Kasieczka, N.~Kiefer, T.~Plehn and J.~M.~Thompson,
SciPost Phys. \textbf{6} (2019) no.6, 069,
[arXiv:1812.09223 [hep-ph]].

\bibitem{Kasieczka:2021xcg}
G.~Kasieczka, B.~Nachman, D.~Shih, O.~Amram, A.~Andreassen, K.~Benkendorfer, B.~Bortolato, G.~Brooijmans, F.~Canelli and J.~H.~Collins, \textit{et al.}
[arXiv:2101.08320 [hep-ph]].

\bibitem{Kasieczka:2020nyd}
G.~Kasieczka, S.~Marzani, G.~Soyez and G.~Stagnitto,
JHEP \textbf{09} (2020), 195,
[arXiv:2007.04319 [hep-ph]].

\bibitem{Coccaro:2019lgs}
A.~Coccaro, M.~Pierini, L.~Silvestrini and R.~Torre,
Eur. Phys. J. C \textbf{80} (2020) no.7, 664,
[arXiv:1911.03305 [hep-ph]].

\bibitem{Radovic:2018dip}
A.~Radovic, M.~Williams, D.~Rousseau, M.~Kagan, D.~Bonacorsi, A.~Himmel, A.~Aurisano, K.~Terao and T.~Wongjirad,
Nature \textbf{560} (2018) no.7716, 41-48,

\bibitem{Guest:2018yhq}
D.~Guest, K.~Cranmer and D.~Whiteson,
Ann. Rev. Nucl. Part. Sci. \textbf{68} (2018), 161-181,
[arXiv:1806.11484 [hep-ex]].

\bibitem{Larkoski:2017jix}
A.~J.~Larkoski, I.~Moult and B.~Nachman,
Phys. Rept. \textbf{841} (2020), 1-63,
[arXiv:1709.04464 [hep-ph]].

\bibitem{Larkoski:2019nwj}
A.~J.~Larkoski and E.~M.~Metodiev,
JHEP \textbf{10} (2019), 014,
[arXiv:1906.01639 [hep-ph]].

\bibitem{Datta:2017rhs}
K.~Datta and A.~Larkoski,
JHEP \textbf{06} (2017), 073,
[arXiv:1704.08249 [hep-ph]].



\bibitem{Aguilar-Saavedra:2020sxp}
J.~A.~Aguilar-Saavedra and B.~Zald\'\i{}var,
Eur. Phys. J. C \textbf{80} (2020) no.6, 530,
[arXiv:2002.12320 [hep-ph]].



\bibitem{Aguilar-Saavedra:2020uhm}
J.~A.~Aguilar-Saavedra, F.~R.~Joaquim and J.~F.~Seabra,
JHEP \textbf{03} (2021), 012,
[arXiv:2008.12792 [hep-ph]].

\bibitem{Komiske:2017aww}
P.~T.~Komiske, E.~M.~Metodiev and J.~Thaler,
JHEP \textbf{04} (2018), 013,
[arXiv:1712.07124 [hep-ph]].

\bibitem{Komiske:2018cqr}
P.~T.~Komiske, E.~M.~Metodiev and J.~Thaler,
JHEP \textbf{01} (2019), 121,
[arXiv:1810.05165 [hep-ph]].

\bibitem{Chakraborty:2020yfc}
A.~Chakraborty, S.~H.~Lim, M.~M.~Nojiri and M.~Takeuchi,
JHEP \textbf{20} (2020), 111,
[arXiv:2003.11787 [hep-ph]].

\bibitem{Fraser:2018ieu}
K.~Fraser and M.~D.~Schwartz,
JHEP \textbf{10} (2018), 093,
[arXiv:1803.08066 [hep-ph]].

\bibitem{Dreyer:2018nbf}
F.~A.~Dreyer, G.~P.~Salam and G.~Soyez,
JHEP \textbf{12} (2018), 064,
[arXiv:1807.04758 [hep-ph]].

\bibitem{deOliveira:2015xxd}
L.~de Oliveira, M.~Kagan, L.~Mackey, B.~Nachman and A.~Schwartzman,
JHEP \textbf{07} (2016), 069,
[arXiv:1511.05190 [hep-ph]].

\bibitem{Komiske:2016rsd}
P.~T.~Komiske, E.~M.~Metodiev and M.~D.~Schwartz,
JHEP \textbf{01} (2017), 110,
[arXiv:1612.01551 [hep-ph]].

\bibitem{Lifson:2020gua}
A.~Lifson, G.~P.~Salam and G.~Soyez,
JHEP \textbf{10} (2020), 170,
[arXiv:2007.06578 [hep-ph]].

\bibitem{Dreyer:2020brq}
F.~A.~Dreyer and H.~Qu,
JHEP \textbf{03} (2021), 052,
[arXiv:2012.08526 [hep-ph]].

\bibitem{Carrazza:2019cnt}
S.~Carrazza and F.~A.~Dreyer,
Eur. Phys. J. C \textbf{79} (2019) no.11, 979,
[arXiv:1909.01359 [hep-ph]].


\bibitem{Dillon:2020quc}
B.~M.~Dillon, D.~A.~Faroughy, J.~F.~Kamenik and M.~Szewc,
JHEP \textbf{10} (2020), 206,
[arXiv:2005.12319 [hep-ph]].

\bibitem{Aad:2020zcn}
G.~Aad \textit{et al.} [ATLAS],
Phys. Rev. Lett. \textbf{124} (2020) no.22, 222002,
[arXiv:2004.03540 [hep-ex]].

\bibitem{ALICE-PUBLIC-2021-002}
ALICE,
https://cds.cern.ch/record/2759456.

\bibitem{CNNKasieczka}
G.~Kasieczka, T.~Plehn, M.~Russell and T.~Schell,
JHEP \textbf{05} (2017), 006,
[arXiv:1701.08784 [hep-ph]].  
  
\bibitem{CNNsShih}
S.~Macaluso and D.~Shih,
JHEP \textbf{10} (2018), 121,
[arXiv:1803.00107 [hep-ph]].  

\bibitem{Khosa:2019qgp}
C.~K.~Khosa, L.~Mars, J.~Richards and V.~Sanz,
J. Phys. G \textbf{47} (2020) no.9, 095201,
[arXiv:1911.09210 [hep-ph]].

\bibitem{Khosa:2019kxd}
C.~K.~Khosa, V.~Sanz and M.~Soughton,
[arXiv:1910.06058 [hep-ph]].

\bibitem{Chung:2020ysf}
Y.~L.~Chung, S.~C.~Hsu and B.~Nachman,
[arXiv:2009.05930 [hep-ph]].

\bibitem{Khosa:2020qrz}
C.~K.~Khosa and V.~Sanz,
[arXiv:2007.14462 [cs.LG]].

\bibitem{Guo:2020vvt}
J.~Guo, J.~Li and T.~Li,
[arXiv:2010.05464 [hep-ph]].

\bibitem{Lin:2018cin}
J.~Lin, M.~Freytsis, I.~Moult and B.~Nachman,
JHEP \textbf{10} (2018), 101,
[arXiv:1807.10768 [hep-ph]].

\bibitem{Li:2020grn}
J.~Li, T.~Li and F.~Z.~Xu,
[arXiv:2008.13529 [hep-ph]].

\bibitem{Alves:2019ppy}
A.~Alves and F.~F.~Freitas,
Mach. Learn. Sci. Tech. \textbf{1} (2020) no.4, 045025,
[arXiv:1912.12532 [hep-ph]].

\bibitem{Moreno:2019neq}
E.~A.~Moreno, T.~Q.~Nguyen, J.~R.~Vlimant, O.~Cerri, H.~B.~Newman, A.~Periwal, M.~Spiropulu, J.~M.~Duarte and M.~Pierini,
Phys. Rev. D \textbf{102} (2020) no.1, 012010,
[arXiv:1909.12285 [hep-ex]].

\bibitem{Datta:2017lxt}
K.~Datta and A.~J.~Larkoski,
JHEP \textbf{03} (2018), 086,
[arXiv:1710.01305 [hep-ph]].

\bibitem{Datta:2019ndh}
K.~Datta, A.~Larkoski and B.~Nachman,
Phys. Rev. D \textbf{100} (2019) no.9, 095016,
[arXiv:1902.07180 [hep-ph]].

\bibitem{Amacker:2020bmn}
J.~Amacker, W.~Balunas, L.~Beresford, D.~Bortoletto, J.~Frost, C.~Issever, J.~Liu, J.~McKee, A.~Micheli and S.~Paredes Saenz, \textit{et al.}
[arXiv:2004.04240 [hep-ph]].

\bibitem{Abdughani:2020xfo}
M.~Abdughani, D.~Wang, L.~Wu, J.~M.~Yang and J.~Zhao,
[arXiv:2005.11086 [hep-ph]].


\bibitem{Ngairangbam:2020ksz}
V.~S.~Ngairangbam, A.~Bhardwaj, P.~Konar and A.~K.~Nayak,
Eur. Phys. J. C \textbf{80} (2020) no.11, 1055,
[arXiv:2008.05434 [hep-ph]].


\bibitem{Harris:2019qwx}
P.~C.~Harris, D.~S.~Rankin and C.~Mantilla Suarez,
[arXiv:1910.02082 [hep-ph]].

\bibitem{Englert:2020ntw}
C.~Englert, M.~Fairbairn, M.~Spannowsky, P.~Stylianou and S.~Varma,
Phys. Rev. D \textbf{102} (2020) no.9, 095027,
[arXiv:2008.08611 [hep-ph]].

\bibitem{Grojean:2020ech}
C.~Grojean, A.~Paul and Z.~Qian,
[arXiv:2011.13945 [hep-ph]].

\bibitem{Buckley:2020kdp}
A.~Buckley, G.~Callea, A.~J.~Larkoski and S.~Marzani,
SciPost Phys. \textbf{9} (2020), 026,
[arXiv:2006.10480 [hep-ph]].

\bibitem{Dokshitzer:1997in}
Y.~L.~Dokshitzer, G.~D.~Leder, S.~Moretti and B.~R.~Webber,
JHEP \textbf{08} (1997), 001,
[arXiv:hep-ph/9707323 [hep-ph]].

\bibitem{Wobisch:1998wt}
M.~Wobisch and T.~Wengler,
[arXiv:hep-ph/9907280 [hep-ph]].

\bibitem{Alwall:2014hca}
J.~Alwall, R.~Frederix, S.~Frixione, V.~Hirschi, F.~Maltoni, O.~Mattelaer, H.~S.~Shao, T.~Stelzer, P.~Torrielli and M.~Zaro,
JHEP \textbf{07} (2014), 079,
[arXiv:1405.0301 [hep-ph]].

\bibitem{Sjostrand:2006za}
T.~Sjostrand, S.~Mrenna and P.~Z.~Skands,
JHEP \textbf{05} (2006), 026,
[arXiv:hep-ph/0603175 [hep-ph]].

\bibitem{Sjostrand:2007gs}
T.~Sjostrand, S.~Mrenna and P.~Z.~Skands,
Comput. Phys. Commun. \textbf{178} (2008), 852-867,
[arXiv:0710.3820 [hep-ph]].


\bibitem{antikt}
M.~Cacciari, G.~P.~Salam and G.~Soyez,
JHEP \textbf{04} (2008), 063,
[arXiv:0802.1189 [hep-ph]].

\bibitem{fastjet}
M.~Cacciari, G.~P.~Salam and G.~Soyez,
Eur. Phys. J. C \textbf{72} (2012), 1896,
[arXiv:1111.6097 [hep-ph]].


\bibitem{keras}
F. Chollet, ``Keras'', https://keras.io, 2015.

\bibitem{Kingma:2014vow}
D.~P.~Kingma and J.~Ba,
[arXiv:1412.6980 [cs.LG]].

\bibitem{heft}
https://feynrules.irmp.ucl.ac.be/wiki/HiggsEffectiveTheory

\bibitem{Dasgupta:2013ihk}
M.~Dasgupta, A.~Fregoso, S.~Marzani and G.~P.~Salam,
JHEP \textbf{09} (2013), 029,
[arXiv:1307.0007 [hep-ph]].

\bibitem{Dasgupta:2013via}
M.~Dasgupta, A.~Fregoso, S.~Marzani and A.~Powling,
Eur. Phys. J. C \textbf{73} (2013) no.11, 2623,
[arXiv:1307.0013 [hep-ph]].

\bibitem{Dasgupta:2015yua}
M.~Dasgupta, A.~Powling and A.~Siodmok,
JHEP \textbf{08} (2015), 079,
[arXiv:1503.01088 [hep-ph]].

\bibitem{Bendavid:2018nar}
J.~R.~Andersen, J.~Bellm, J.~Bendavid, N.~Berger, D.~Bhatia, B.~Biedermann, S.~Br\"auer, D.~Britzger, A.~G.~Buckley and R.~Camacho, \textit{et al.}
[arXiv:1803.07977 [hep-ph]].


\end{thebibliography}
\end{document}